\documentstyle[12pt]{article}                                                  
\newcommand{\beq}{\begin{equation}}
\newcommand{\eeq}{\end{equation}}
\newcommand{\beqn}{\begin{eqnarray}}
\newcommand{\eeqn}{\end{eqnarray}}
\def\ppdir{p'\kern-7.8pt\Big{/}}
\def\pdir{p\kern-7.8pt\Big{/}}
\def\kdir{k\kern-7.8pt\Big{/}}
\def\qdir{q\kern-7.8pt\Big{/}}

\begin{document}  

\title{Introduction to Perturbative QCD}
\author{ 
{\it U.~Aglietti}\\
~~~~~~~~~~~~~~~~~\\ 
Dipartimento di Fisica, Universit\` a di Roma \lq La Sapienza \rq \\ 
INFN, Sezione di Roma I, P.le A. Moro 2, 00185 Roma, Italy \\}
%\mbox{} \\ 
\date{}
\maketitle
\begin{abstract}
\noindent 
This is a written version of two lectures given at the
First School on Field Theory and Gravitation
in Vitoria (Brasil) April 15-19, 1997.
We discuss the foundation of QCD as the theory of strong 
interactions and the perturbative analysis
of $e^+e^-$ annihilation to hadrons.
Typical concepts of perturbative QCD studies, such as collinear
singularities, jets, Sudakov form factors, are explained
working out this case.
\end{abstract}

\newpage

\section{Introduction}

Quantum Chromodynamics (QCD), introduced by Gell-Mann and Frizsch in
1972 \cite{frigel}, is the current theory of strong interactions.
It is a renormalizable nonabelian gauge theory 
\cite{nonabel} based on the group 
$SU(3)$, containg quark and gluons as elementary fields, 
\beq
L~=~-\frac{1}{4}F_{\mu\nu}^aF^{\mu\nu a}
+\sum_f\overline{q}_f(i\gamma_{\mu}D^{\mu}-m_f)q_f
\label{eq:qcd}\eeq
where $F_{\mu\nu}^a=\partial_{\mu}A_{\nu}^a-\partial_{\nu}A_{\mu}^a
-gf^{abc}A_{\mu}^bA_{\nu}^c$
is the field tensor, $D_{\mu}=\partial_{\mu}+igA_{\mu}^at^a$ is the
covariant derivative, $t^a$ are the Gell-Mann matrices, $f^{abc}$
are the structure constants of $SU(3)$ and $f$ denotes a quark flavor.

Strong interactions present many different phenomena  
and QCD at present cannot account for all of them.
To give an example, QCD is irrelevant to 
the entire field of low energy nuclear physics.
This situation is to be contrasted with QED, where most
processes can be computed with high accuracy.

In sec. \ref{found} we discuss the foundation of QCD, i.e. why we
believe we have the correct theory 
(the lagrangian (\ref{eq:qcd})) even though many
phenomena are not described or computed inside it.

In sec. \ref{asfr} we discuss the fundamental property of QCD,
asymptotic freedom, according to which QCD approaches a
{\it free} theory in the ultraviolet region.
This property explains parton model assumptions
and allows perturbative computations as corrections
to the free quark behaviour.

In sec. \ref{conf} a qualitative discussion is presented of quark 
confinement,
according to which quarks do not exist as isolated particles,
but only in bound states, the observed hadrons.
This phenomenon cannot be 
described {\it by definition} in perturbative QCD,
which assumes that asymptotic states, the hadrons, are 
infinitesimally close to the free ones, quarks and gluons.
There are indications of confinement from lattice
(non perturbative) QCD computations, but a rigorous  proof 
is missing. This section describes
some qualitative ideas about confinement, 
to understand why perturbative QCD 
can be applied {\it despite} confinement. 
We discuss also the production of hadronic jets,
a typical phenomeon of high energy hadronic processes.
The relation with confinement is explained, and also why
we believe that a perturbative QCD computation can describe
seminclusive jet properties. 

In sec. \ref{eehad} we consider the production of hadrons
in $e^+e^-$ annihilation at high energy as a typical
application of perturbative QCD.
This is a simpler process than 
lepton-hadron or hadron-hadron scattering, 
so it is a good starting point for introducing the reader to
perturbative QCD.
The total hadronic cross section is 
perhaps the easiest thing to compute.
Furthermore, the assumptions on hadronization 
to apply perturbative QCD are minimal:
we require only that hadronization does not change the 
probability of the partonic process.

Later we study the structure of the final hadronic
state, which reveals many typical perturbative QCD phenomena,
like collinear singularities, jets, 
infrared effects to be factorized, etc. 
In the latter case a stronger assumption 
about hadronization has to be made:
the momentum flow of the partonic event has to remain
substantially the same after hadronization.  

There is also a section with the conclusions and an appendix
describing the computation of $e^+e^-$ annihilation into
three jets.

\section{Foundation}
\label{found}

QCD is a non abelian gauge theory based on the group $SU(3)$
of color.
The main motivation for a non abelian gauge theory 
came from the parton model of deep
inelastic scattering, according to which
quarks behave like free particles in hard collisions.
This model has been successively extended,
with similar success,
to describe also hard hadron-hadron collisions 
and hadronic production in $e^+e^-$ annihilation
at high energy (sec. \ref{eehad}).
The hystory of deep inelastic
scattering and the ideas of the parton model are 
sketched in sec. \ref{pm}.
 
The choice of the gauge group and of the quark representation
is motivated by various phenomena which are
discussed in sec. \ref{col}.

We believe that all these different phenomena, collectively,
give a good support to the structure of QCD as the theory
of strong interactions.

\subsection{The Parton Model}
\label{pm}

A series of experiments were performed at SLAC
in the sixties 
to understand the structure of the proton and of the neutron,
measuring the so called electric and magnetic form factors.
Electrons with an energy up to 20 GeV were sent against a target
of hydrogen or deuterium $N$
\beq
e+N\rightarrow e+X
\eeq
where $X$ is any hadronic final state.

The astonishing result was that a larger number than expected 
of large angle deflections of the electron were observed.
Feynman gave a simple phenomenological explanation for this
result:
the nucleon has to be considered in deep inelastic collisions
as a gas of non interacting pointlike 
particles, the partons;
the electron simply suffers an elastic collision with a parton $p$
\beq
e+p\rightarrow e+p
\eeq
A pointlike cross section has not the
form factor suppression of an extended object.
We have therefore 'hard' interactions, with large
angle deflections of the electron, as observed.
To have a physical picture, consider the nucleon in a frame 
in which it has a relativistic velocity $(\gamma_{Lorentz}\gg 1$),
for example the rest frame of the colliding electron.
Due to the Lorentz boost, the nucleon looks like a bunch of
collinear partons, each carrying a fraction $x_q$ of the
nucleon momentum, such that
\beq
\sum_q x_q~=~1.
\label{eq:sr}\eeq
A hadron is represented by a function (partonic density)
$q(x)$ telling how many partons $dn$ there are with
momentum fraction $x$:
\beq
dn(x\div x+dx)~=~q(x)dx
\eeq
The observed hadronic cross section is the convolution
of the parton density with the pointlike, partonic
cross section.
\beq
\sigma_{hadr}~=~q*\sigma_{part}
\eeq
Since the pointlike cross section assumes free partons,
all the dynamical effects of strong interactions
are contained in the specific form of $q(x)$.
We see here the idea of factorization at work:
different dynamical processes are represented
by separated factors in the cross section.
This idea has been gathered and generalized
by perturbative QCD, in many different forms:
factorization of mass singularites, Operator
Product Expansion \cite{wilson} (for a detailed discussion see for
example \cite{marti}), Sudakov form factors (see sec. \ref{sud} 
and ref. \cite{collins}), etc..

Parton model was also able to explain a phenomenological law,
the so called Bjorken scaling, i.e the independence of the
form factors on $Q^2$, which coincide
just with the parton densities.

These partons were identified with the quarks introduced 
to classify the variety of hadrons discovered and their spectroscopy.
Partons were assigned a spin $s=1/2$, an electric charge
$e=2/3,-1/3$, a flavor, etc.
That gave rise to a series of sum rules relating
parton densities of different hadrons, 
which were experimentally satisfied.

This identification between 'low energy' and 'high energy'
entities came out to be non trivial.
The valence quarks, the degrees of freedom of the quark model,
were not sufficient to account for momentum conservation
in $e-N$ collisions. The sum rule (\ref{eq:sr}) seemed to be violated,
the right hand side being substantially less than one.
It was necessary to postulate the existence of 'sea' quarks,
in addition to the valence quarks,
short living quark-antiquark pairs in the nucleon,
partecipating to the hard processes.
This idea was quite reasonable from the quantum field theory point
of view, dealing with virtual particles form the very beginning,
but was still incomplete. Half of the momentum of the nucleons 
was not carried neither by the valence quarks nor by the sea quarks
because:
\beq
\sum_{val} x+\sum_{sea} x ~\simeq~0.5
\eeq
It was necessary to admit the existence of neutral particles
in the nucleon
not interacting with the probe (the $\gamma,~W$ or $Z$),
i.e. without electric or weak charge.
These components were naturally identified with the gluons
after the rise of QCD.

\subsection{The Color}
\label{col}

Nonabelian gauge theories are a good candidate to describe
strong interactions because of asymptotic freedom
(they are indeed a {\it unique} candidate in the
framework of relativistic local field theories).
The choice of the gauge group is motivated by a variety
of phenomena, some of which are listed below (see also
\cite{marti} and the lectures of B. Mele at this School
for a more recent compilation). 
These phenomena {\it per se} require only the color group
as an (exact) global symmetry.
Color enters as a multiplicity factor in the
cross section (counting of degrees
of freedom) or as a quantum number 
to distinguish quarks.

\subsubsection{Spectroscopy}

Consider the $\Delta^{++}$ in the framework of a constituent
quark model.
It is a spin $3/2$ baryon and is composed of three
identical $up$ quarks. Since it is the lowest lying state with these
quantum numbers, it is natural to assume that the spatial
wave function is the fundamental one.
A 3-body potential which gives a reasonable description of the
lowest excitations is the harmonic one:
\beq
V(x_1,x_2,x_3)~=~\frac{m\omega^2}{6}( (x_1-x_2)^2+(x_1-x_3)^2+(x_2-x_3)^2)
\eeq
where $x_i$ are the quark coordinates, $m$ is the constituent $up$
quark mass and $\omega$ is a constant to be
determined experimentally.

\noindent
The ground state of the model defined by the Hamiltonian
\beq
H~=~-\frac{1}{2m}\sum_{i=1}^3\frac{\partial^2}{\partial x_i^2}+
V(x_1,x_2,x_3)
\eeq
is a gaussian,
\beq
\psi_0(x_1,x_2,x_3)=\frac{1}{\sqrt{3\sqrt{3}\pi^2}R^3}
\exp\Biggl[ -\Big((x_1-x_2)^2+(x_1-x_3)^2+(x_2-x_3)^2\Big)/(6R^2) \Biggr]
\eeq
where $R^2=1/(m\omega)$.
$\psi_0$ is symmetric under exchange of any pair of coordinates
and it has zero angular momentum (it is invariant under rotations):
\beq
L~=~0.
\eeq
The spin of the baryon 
\beq
\vec{J}~=~\vec{L}+\vec{S}
\eeq
is enterely carried by the
total spin of the quarks $S$:
\beq
J~=~S.
\eeq
The state with $S_z=3/2$ 
consists of the {\it up} quarks all in the positive $z$ direction
\beq
\mid \Delta^{++}, S_z=3/2\rangle~=~\mid u \uparrow, u\uparrow, 
u\uparrow\rangle
\eeq
The spin wave function therefore is also
symmetrical under exchange of the spins, so the complete
wavefunction is symmetrical under exchange of the quarks.
This is in contradiction with the spin-statistics theorem
(the old Pauli principle of atomic physics)
according to which identical spin 1/2 particles must have an 
antisymmetric wavefunction.
A natural solution of this problem is the introduction
of an additional quantum number of the quarks, called the
'color'. There have to be at least three colors.
The most economical solution is that
quarks come in just three varieties of color,
and the baryon wavefunction is antisimetrical under
exchange of the color of any two quarks, so it is a color
singlet.
In color space the wavefunction is therefore
\beq
\mid \Delta^{++}\rangle~=~\frac{1}{\sqrt{6}}\epsilon_{abc}
\mid u_a,u_b,u_c\rangle,
\eeq
where $\epsilon_{abc}$ is the totally antisymmetric tensor
with $\epsilon_{123}=1$.

Similar problems exist also
for the statistics of the $\Delta^-$ and $\Omega^-$ 
baryons,
which are composed of three identical $d$ (down) and $s$ (strange)
quarks respectively.
Assuming $SU(3)$ flavor symmetry, the same problem holds for many
light baryons and is solved in the same way by the color.

Mesons consist of quark antiquark pairs with the following
color singlet wavefunction
\beq
\mid M\rangle~=~\frac{1}{\sqrt{3}}\delta^a_b 
\mid \overline{q}_a~{q'}^b\rangle
\eeq 
where $q$ and $q'$ denote two arbitrary quark flavors.

\subsubsection{$\pi^0\rightarrow \gamma\gamma$ decay}

The $\pi^0$ has the following valence quark content:
\beq
\mid \pi^0 \rangle ~=~\frac{1}{\sqrt{2}}
(\mid u\overline{u}\rangle -\mid d\overline{d} \rangle)
\eeq
We may take as an interpolating field for the $\pi^0$
(an interpolating field is an operator which excites the particle acting on 
the vacuum) the divergence of the isospin $I=1,I_3=0$ axial current
\beq
A_{\mu}~=~\frac{1}{\sqrt{2}}(
\overline{u}\gamma_{\mu}\gamma_5 u -
\overline{d}\gamma_{\mu}\gamma_5 d)
\eeq
Due to the triangle anomaly, there is a non zero coupling
of the axial current with two electromagnetic currents,
\beq
\langle 0\mid T A_{\mu}(x) J_{\nu}(y) J_{\rho}(z) \mid 0\rangle~\neq~0
\eeq
which produces the decay of the $\pi^0$ into two photons.
The computation gives
\beq
\Gamma~=~ \frac{\alpha^2}{64\pi^3} \frac{m_{\pi}^3}{f_{\pi}^2}\xi^2
~=~7.6~\xi^2~eV,
\eeq
where $f_{\pi}$ is a contant determined from leptonic
charged $\pi$ decay (assuming isospin symmetry) and
\beq
\xi~=~N(e_u^2-e_d^2)~=~1
\eeq
with the standard charge assignements and $N=3$.

The decay amplitude is proportional to 
$N$ because the pion couples to $q\overline{q}$ pairs of any color.

\noindent
The measured value is
\beq
\Gamma_{mis}~=~7.7 \pm 0.6~eV
\eeq
in good agrement with the $N=3$ rate.
Note that this case is very sensitive to the number of colors,
due to the quadratic dependence of the rate on $N$.

\subsubsection{Anomaly Cancellation In The Standard Model}

A pretty theoretical argument for the color 
(i.e. not based on any experiment) is the anomaly
cancellation in the Standard Model.
Anomaly has to cancel because
a non vanishing anomaly induces terms of canonical dimension
in the lagrangian
which are not originally present and are therefore
not compatible with gauge invariance. The breaking of 
gauge invariance renders the theory not renormalizable,
destroying the simplicity of the model (the simple
ultraviolet structure). 

No triangle anomaly is generated if the sum 
of the charges for every generation is zero. Since
\beq
e_u+e_d~=~1/3
\eeq
while
\beq
e_{\nu}+e_{l}~=~-1,
\eeq
a multiplicity for the quark states of 3 is required,
which is provided by the color.

\section{Asymptotic Freedom}
\label{asfr}

After the success of the parton model, 
the theoretical problem was that of 
recoinciling a good
phenomenological model with field theory (so succesful in the
area of electromagnetic and weak interactions).
That appeared to be a hard task.
Why was a quark behaving as a free particle when involved in an
energetic collision? What about his interactions,
which actually give rise to his binding in the nucleon?

The resolution of this problem came with the renormalization
of nonabelian gauge theories. In such theories 
the one-loop $\beta$-function is negative.
The $\beta$-function is defined as
\beq
\beta(\alpha_S)~=~\mu^2\frac{\partial \alpha_S}{\partial\mu^2}
\label{eq:eqdif}\eeq
and represents the variation of the renormalized coupling
as we vary the renormalization point $\mu$ 
(see also sec. \ref{radcor} to understand its physical meaning).
In this variation we have to keep the bare couplings 
($\alpha_S^{(0)}$ and the bare masses $m_0$'s)
and the ultraviolet cutoff $\Lambda_0^2$ fixed to ensure that we
remain within the same physical theory (parametrized in 
many different ways according to the many different choices of the scale 
$\mu$ to which the parameters of the theory can be defined).
Alternatively, we may define the $\beta$-function as
\beq
\beta(\alpha_{S}^0)~=~\Lambda_0^2
\frac{\partial\alpha_{S}^0}{\partial\Lambda_0^2},
\eeq
i.e. the variation of the bare coupling as we vary the 
ultraviolet cutoff $\Lambda_0$.
The variation is done keeping the renormalized parameters
constant, so as to remain in
the same physical theory, i.e. to reproduce the same
low-energy cross sections (with energy
$s\ll\Lambda_0^2$). 
It can be shown that the two above definitions give the
same function up to two loops \cite{alt2}, beyond
which details of the regularization (the precise way
we cut the ultraviolet region) produce
differences in the coefficients.
The $\beta$-function has a perturbative expansion
which we write in the form
\beqn
\frac{\beta(\alpha_S)}{4\pi}&=&\sum_{n=0}^{\infty}
     \beta_n\Biggl(\frac{\alpha_S}{4\pi}\Biggr)^{n+2}
\nonumber\\
&=&\beta_0\Biggl(\frac{\alpha_S}{4\pi}\Biggr)^2
+\beta_1\Biggl(\frac{\alpha_S}{4\pi}\Biggr)^3
+\beta_2\Biggl(\frac{\alpha_S}{4\pi}\Biggr)^4+...
\eeqn
We divided $\beta$ by $4\pi$ because
the one-loop beta function is already $O(\alpha_S^2)$.

\noindent
The one-loop value is
\beqn
\beta_0&=&-\Bigl(\frac{11}{3}N-\frac{2}{3}n_f\Bigr)
\nonumber\\
&=&-11+\frac{2}{3}n_f~=~-7.67~~~~~~~(for~n_f=5)
\eeqn
where $N=3$, is the number of colors
(as proved in the previous section),
and $n_f$ is the number of 'active' flavors at the 
scale $\mu$.
With $f<11 N/2=16.5$ flavors, i.e. $f\leq 16$,
\beq
\beta_0~<0
\eeq 
The differential equation (\ref{eq:eqdif}) can be integrated 
approximating the $\beta$-function
with its one-loop value.
Denoting the initial integration condition with the scale $\mu^2$
and the final one with the scale $Q^2$, we have

\beq
\alpha_S(Q^2)~=~\frac{\alpha_S(\mu^2)}
               {1-\beta_0/(4\pi)\alpha_S(\mu^2)\log(Q^2/\mu^2)}
\label{eq:dopo}
\eeq

We see that, since $\beta_0<0$, the coupling
$\alpha_S(Q^2)$ decreases with the energy and approaches
zero in the limit of a a very large energy
(or momentum transfer):
\beq
\alpha_S(Q^2)\rightarrow 0~~~~~~~~{\rm for}~~~~~Q^2\rightarrow \infty
\eeq
The theory therefore approaches a free theory in the ultraviolet
region (asymptotic freedom \cite{asfr}),
thus explaining the 'free' behaviour of the quark in the
hard process.
On the other side, the coupling is not small at low energies,
explaining quark binding in the hadrons.
We have therefore a very simple explanation of the
duality of high-energy and low-energy hadronic
phenomena with renormalization group ideas.
We note that asymptotic freedom is a one-loop property
of the $\beta$-function because at large enough energies
the coupling is small and higher orders are negligible.

Instead of introducing a reference scale $\mu^2$ at which
assign the value of the coupling, we can write
the running coupling in terms of 
the so-called $\Lambda_{QCD}$ parameter, the scale
at which the one-loop coupling has a pole
(it is an infrared pole because it occurs at low energies).
With
\beq
\Lambda_{QCD}^2~=~\mu^2e^{4\pi/(\beta_0\alpha_S(\mu^2))}
\eeq 
the coupling is rewritten as
\beq
\alpha_S(Q^2)~=~\frac{1}{-\beta_0/(4\pi)\log(Q^2/\Lambda_{QCD}^2)}.
\label{eq:asfr}\eeq
We may ask what is the physical meaning of $\Lambda_{QCD}$.
It is a renormalization group invariant quantity
(it does not change
if we vary $\mu^2$ in its defining equation)
and denotes at which energy renormalization group
improved perturbation theory ceases to be valid, pointing 
therefore to generic nonperturbative phenomena \cite{me}.
This scale has to be determined experimentally and turns
out to be $\sim 300~MeV$, i.e. of the same order of the
hadron masses ($M_{\rho} \sim 770~MeV$).

We note that the coupling goes to zero only logarithmically
with the energy, i.e. very slowly.
As long as $\alpha_s\ll 1$, perturbative corrections 
to the free theory behaviour can be computed and are expected to be
sizable.
This is the technical justification for perturbative QCD, whose
physical justification will be discussed later.

Let us quote also the values of two and three loop coefficients of the
$\beta$-function in the $\overline{MS}$ scheme 
\cite{beta3l}:
\beqn
\beta_1&=&-102+\frac{38}{3}n_f
~=~-102+12.66~n_f~=~-38.66
\nonumber\\
\beta_2&=&\frac{1}{2}
\Big[-2857+\frac{5033}{9}n_f-\frac{325}{27}n_f^2\Big]
\nonumber\\
&=&-1428.5+279.6n_f-6.0n_f^2~=~-180.5~~~~~~~~~(for~n_f=5)
\eeqn
We note that $\beta_1<0$ if $n_f\leq 8$, while $\beta_2<0$ if
$n_f\leq 5$.
The (approximate) two-loop solution for the running
coupling is given by:
\beq
\alpha_S(Q^2)~=~\frac{1}{-\beta_0/(4\pi)\log(Q^2/\Lambda_{QCD}^2)}
\Bigg[1+\frac{\beta_1}{\beta_0^2}
\frac{\log\Bigl(\log(Q^2/\Lambda^2)\Bigr)}
     {\log (Q^2/\Lambda^2)}\Bigg].
\eeq
In QED the $\beta$-function is given to one loop by
\beq
\beta_{QED}=\frac{\alpha^2}{3\pi}~>~0
\eeq
(times $n_f$ if $f$ flavors are considered).
Integrating the related differential equation exactly
as in the case of $QCD$, we derive:
\beq
\alpha(Q^2)~=~\frac{3\pi}{\log (\Lambda_L^2/Q^2)}
\eeq
where $\Lambda_L$, called the Landau pole, is now an
ultraviolet scale \cite{landau}.
The coupling grows with the energy as a consequence of
the positive $\beta$-function.
The QED behaviour of the effective coupling is therefore opposite 
to that of QCD (eq.(\ref{eq:asfr})). 
A physical discussion of these different behaviours
is given in the next section.

We conclude this section observing that 
in four space-time dimensions  asymptotic freedom is an 
exclusive property of non abelian gauge theories \cite{cl}, which therefore 
emerge as a natural candidate for the strong interactions.

\subsection{Antiscreening}

In QED the behaviour of the effective charge is controlled by
the photon polarization diagram,
\beq
\gamma^*~\rightarrow~(e^+e^-)^*~\rightarrow~\gamma^*
\eeq
(ultraviolet divergences
in the electron field renormalization and in the vertex
correction cancel because of the Ward identity).

The physical explanation is the following.
Consider a charged particle
placed in the vacuum, with bare (i.e. primordial) charge $e_0$. 
Quantum field fluctuations induce short lived $e^+e^-$ pairs.
These pairs constitute electric dipoles which align
with the test charge so as to minimize the energy.
The bare charge is therefore surrounded by a shell of
opposite charges (the outer shell is at infinity)
with a consequent screening effect.
The vacuum behaves as an ordinary medium
with material electric dipoles, with dielectric constant 
\beq
\epsilon~>~1~~~~~~~~~~~(QED)
\eeq
If we measure the charge of the test particle, we find that it
is less than the bare charge:
\beq
e~=~\frac{e_0}{\epsilon}
\eeq
If we come closer to the test particle, say at distance
$r$, we see an increasing
charge, because the outer part of the shell, at distance
$r'>r$, does not
contribute to the screening (Gauss theorem).
The effective charge (i.e. the observable one) is 
therefore a function of the distance 
\beq
e~=~e(r).
\eeq
It increases as we lower the distance, or equivalently, as we
increase the momentum transfer $q^2\sim 1/r^2$.

In QCD the situation is reversed: due to gluon-gluon couplings
(three and four gluon vertices),
the test charge is surrounded by gluons of the {\it same } charge.
The vucuum behaves as an hypotetical medium with dielectric 
constant 
\beq
\epsilon~<~1~~~~~~~~~~~~~~~~(QCD)
\eeq
so we
have an antiscreening effect. As we come closer to the test particle,
the effective charge is reduced (see \cite{kell,gw} for a 
more detailed explanation).
Asymptotic freedom means that the measured charge approaches
zero at an infinitesimal distance.

\section{Confinement}
\label{conf}

Isolated quarks have never been observed, as particles with fractional
electric charge, for example. 
Hence the hypotesis that they are
'confined' in hadrons, i.e. that they do not appear as asymptotic states
in any physical process.
The current explanation is that the force between two quarks
does not vanish as their separation $r$ increases, so that
the potential energy of the system $V$ diverges 
\beq
V(r)~\rightarrow~\infty~~~~~as~~~~~r~\rightarrow~\infty.
\eeq
In any realizable collision, the available energy is finite,
making therefore impossible to produce isolated quarks. 

A potential which is belived to represent the QCD interaction
between a quark and an antiquark in a color singlet state,
is the so called 'funnel' potential: 
\beq
V(r)~=~-\frac{4}{3}\frac{\alpha_s}{r}+kr
\label{eq:funnel}\eeq
where $k$ is a constant, called the string tension.

\noindent
This potential has been extensively used in non relativistic models
of $c\overline{c}$ and $b\overline{b}$ bound states, with 
good results, and is also compatible with lattice 
(non perturbative) QCD computations. 
We see that at small distances, $r\ll 1/k$, the potential
is coulombic, as it is in perturbative QCD, while
at large distances there is a linear grow of the energy with
the separation. 

The  qualitative explanation of the confining potential
(\ref{eq:funnel}) is the following.
When the separation of the quark-antiquark pair
is small compared to the confinement radius,
the gluon field has the same form of a dipole field in
classical electrodynamics. Chromoelectric field lines
spread out in all the space (and the potential decays as $1/r$
like in QED).
When their separation becomes of the 
order of the confinement
radius, there is a reciprocal attraction of the field lines,
which makes them to collapse in a tube 
along the $q\overline{q}$ line.  
Since the electric flux is the same through any closed surphace
containing the quark or the antiquark, 
and the electric field is non zero only inside the tube,
we say that there is
the formation of a flux tube (or string).
A further increase of the separation does not change the form of
the gluon field anymore but only makes the flux tube longer.
Since the gluon field is now essentially concentrated in a tube,
the energy contained in the tube is proportional to its lenght.
In other words, there is a constant energy of the field
per unit lenght, which implies a linear term in the potential.
In the next section we discuss a model of confinement
which induces the above postulated attraction of field lines.

\subsection{Dual Meissner effect}

A model which induces flux tubes (and therefore
confinement via a linear potential) 
is the so called dual Meissner effect \cite{ash}.

Consider a superconductor 
in an external magnetic field $\vec{B}$ 
above the transition temperature $T_C$,
for definiteness a cilinder with its axis parallel to $\vec{B}$.
As we lower the temperature to $T<T_C$, there is the transition from
the normal to the superconducting phase (drop to zero of the
electrical resistence). At the same time, the magnetic field
lines are expelled outside the sample. 
The magnetic field inside the cilinder is zero:
\beq
\vec{B}_{inside}=~0,
\eeq
i.e. the superconductor shows a perfect diamagnetism
\beq
\chi _M~=~0.
\eeq
The explanation is that with the transition
electrons pair in particles with charge $2e$ (Cooper pairs). 
The latter induce a current $\vec{J}_{pairs}$ 
along the surphace of the cilinder generating 
a magnetic field inside the sample opposite to the external one:
\beq
\vec{B}_{pairs}~=~-\vec{B}
\eeq
Now imagine of placing a pair of opposite charge magnetic monopoles 
inside the superconductor. 
Since the superconductor tends to
expel the magnetic field, the field lines of the
magnetic dipole are shrunk in a small tube connecting the
monopoles. 
An electric current of Cooper pairs 
develops along the surphace of the flux tube,
effectively squeezing the magnetic dipole field.

As well know, electromagnetism is invariant under the 
simultaneous exchange
of magnetic fields with electric ones and of magnetic charges with
electric ones (duality symmetry).

Imagine the dual of an ordinary superconductor,
i.e. a system expelling electric fields.
If we place a pair of opposite electric charges inside
the system, the lines of the electric dipole field 
will be shrunk into an electric flux tube.
The dual superconductor contains magnetic monopoles, 
while the ordinary superconductor contains  
'Cooper pairs, i.e. 'electric monopoles'.
Let us describe the phenomenon a little bit more in detail.
Soon after the introduction of the electric charges, 
the system reacts producing a current of magnetic monopoles
$\vec{J}_m$ along the surphace of a tube connecting the test charges. 
$\vec{J}_m$ produces, by duality,
a secondary electric field $\vec{E}_m$ cancelling
the external field $\vec{E}$ outside the tube
\beq
\vec{E}_m~=-\vec{E}~~~~~~~(outside)
\eeq
and reinforcing the external field inside the tube.
The net effect is the squeezing of the 
dipole electric field of the pair into a tube.

Confinement has not been rigorously proved in QCD up to now.
It is usually assumed that
confinement means that hadrons consist of color singlets only.

\subsection{Jets}
\label{jets}

The process of jet formation in high energy collisions
is qualitatively related to confinement.
As for confinement, there is no rigorous proof of this
phenomenon and of the ideas explaining it inside QCD.

Let us consider the simplest case of jet production,
$e^+e^-$ collisions, to be discussed in detail in the next
section.
The electron and the positron annihilate into a photon, 
which decays into a quark-antiquark pair:
\beq
e^++e^-~\rightarrow~\gamma~\rightarrow q+\overline{q}.
\eeq 
In the center of mass frame,
the quark and the antiquark have
the same energy and opposite spatial momenta
(back to back).
They are originally created in the same spatial point 
(the fireball) and fly far apart 
with light velocity. 
When their separation comes close to the confinement radius, 
\beq
r_{conf}~\sim~ 1~fm~=~10^{-13}~cm,
\eeq
the dipole color field of the quark pair shrinks to a tube
(called flux tube of string).
As they separate from each other, the flux tube gets longer and 
an increasing fraction of the kinetic energy of the pair is 
converted into field energy.
When the energy contained in the flux tube exceeds the mass of a light
$q\overline{q}$ pair
\beq
E_{tube}~> 2m,
\eeq 
a novel quark-antiquark pair is created which screens the color
of the original pair. The string breaks into two shorter strings,
each having at his ends an original quark and a novel quark
in a color singlet state.  
  
If the relative momentum of the quark-antiquark pair connected
to the same string is large enough, the string may break again.
We have therefore the production of a parton cascade, i.e. of
a large number of quarks and gluons with progressively lower energy.

In high energy processes ($E\gg M_P$, say, where $M_P$ is the
proton mass), 
the flux tube has a  large longitudinal momentum ($P_L\sim E$), while it 
has a limited tranverse momentum ($P_T\sim M_P$). Consequently, the 
partons produced have large longitudinal momenta and
large longitudinal momentum differences, but limited 
transverse momenta. 

After the parton cascade production, we have the recombination
of quark and gluons in color singlet states, the observed
hadrons. 
This transition does not induce large momentum
transfer between the partons (it is a soft process). 
This hypotesis is
sometimes called local parton-hadron duality and states
that hadrons are produced by partons which are close
in phase space \cite{nason}. 
This mechanism forbids, for example, that a quark 
with momentum $p$ coming from one jet combines with a quark 
with momentum $p'$ coming from another jet 
($p\cdot p'\gg\Lambda_{QCD}^2$), destroying
the shape of the perturbative momentum distribution.
The transition from a partonic jet to a hadronic one
therefore does not wash out all the parton information and
the longitudinal and transverse momentum
distributions are substantially unchanged.
The final particles, the hadrons, are collimated into two small 
angular regions (jets) opposite to each other see (fig.1).

Furthermore, the angular distribution of the jets relative
to the $e^+e^-$ flight direction is expected to coincide
with that one of the original $q\overline{q}$ pair.
As we will see, this prediction is confirmed by data,
giving good support to the above qualitative ideas.

If an energetic gluon is produced at large angle
with respect to the original pair (a hard gluon), a new 
string is formed with a consequent third jet (see fig.2).

From these considerations, it emerges that jets are a 
universal occurence of high energy hadronic reactions.

\section{Hadron production in $e^+e^-$ collisions}
\label{eehad}

Another evidence for the color comes from the hadronic
production in $e^+e^-$ machines
(in this section we rest on ref. \cite{kell}).
Consider the production of a fermion pair
In lowest order, the electron and the positron annihilate into a 
$\gamma$ or a $Z^0$, which decay into a fermion-antifermion pair
($f\neq e$), (see fig. 3)
\beq
e^++e^-\rightarrow \gamma^*,Z^*\rightarrow f+\overline{f}
\eeq

The differential cross section is given by:
\beqn
\frac{d\sigma}{d\cos\theta}&=&\frac{\pi\alpha^2}{2}\frac{1}{s}
\Biggl[ (1+\cos^2\theta)\Biggl(e_f^2-2e_fv_ev_f\chi_1(s)+
(a_e^2+v_e^2)(a_f^2+v_f^2)\chi_2(s)\Biggr)
\nonumber\\
&+&4\cos\theta\Biggl(2a_ev_ea_fv_f\chi_2(s)
-e_fa_ea_f\chi_1(s)\Biggr)\Biggr] 
\label{eq:basic}\eeqn
where $\theta$ is the angle between the initial and the final
pair,
\beqn
\chi_1(s)&=&k\frac{s(s-M_Z^2)}{(s-M_Z^2)^2+\Gamma_Z^2M_Z^2},
\nonumber\\
\chi_2(s)&=&k^2\frac{s^2}{(s-M_Z^2)^2+\Gamma_Z^2 M_Z^2},
\\
k&=&\frac{\sqrt{2}G_F M_Z^2}{4\pi\alpha}~=~1.40
\eeqn
($k$ is the adimensional ratio of the relevant weak coupling to the 
electromagnetic one and its numerical value is for 
$\alpha(M_Z)=1/128$ \cite{pdg}).
$G_F$ is the Fermi constant, $\alpha$ is the fine structure constant,
$M_Z$ and $\Gamma_Z$ are the mass and the width of the $Z$ boson
respectively.
$v_f$ and $a_f$ are the vector and axial vector couplings of the
$Z$ to the fermions and are given by:
\beqn
v_f&=&I_{3f}-2e_f\sin^2\theta_W
\nonumber\\
a_f&=&I_{3f}
\label{eq:isospin}\eeqn
where $\theta_W$ is the Weinberg angle, and $I_3$ is the 
$z$-component of the weak isospin:
$I_3=1/2$ for neutrinos and {\it u}-type quarks 
($e=2/3$), and $I_3=-1/2$ 
for charged leptons and {\it d}-type quarks ($e=-1/3$).

The function $\chi_1(s)$ represents the interference of the 
$\gamma$ amplitude with the $Z$ one, and the function $\chi_2(s)$ 
represents the $Z$ amplitude squared.

At small energies compared to the $Z$ mass
\beq
s~\ll~M_Z^2
\eeq
the functions $\chi_1$ and $\chi_2$ are suppressed by the $Z$ mass:
\beq
\chi_1\sim -k \frac{s}{M_Z^2} \ll 1,~~~~~~
\chi_2\sim\chi_1^2 \ll 1
\eeq
We can therefore neglect both the direct $Z$ effect 
and the interference one ($\chi_1=\chi_2=0$), 
ending up with the pure QED cross section
\beq
\frac{d\sigma}{d\cos\theta}~=~\frac{\pi\alpha^2}{2}~e_f^2~\frac{1}{s}
(1+\cos^2\theta)
\eeq
The angular dependence is in good agreement with the observed
2-jet cross section and gives support both to the spin=1/2 
assignement of the quarks and to the qualitative ideas
about jet formation discussed previously (see fig.4).

LEP1 and SLAC machines have operated on the $Z$ peak, i.e.
with a center of mass energy $E=M_Z$.
In this case the resonant contribution of the $Z$ dominates
over the $\gamma$ and the interference ones.
Setting therefore $e_f=0,~\chi_1=0$ and $\chi_2=k^2 M_Z^2/\Gamma_Z^2$
we have for the differential cross section
\beq
\frac{d\sigma}{d\cos\theta}~=~\frac{\pi\alpha^2 k^2}{2\Gamma_Z^2}
\Biggl[ (1+\cos^2\theta)
(a_e^2+v_e^2)(a_f^2+v_f^2)
+8\cos\theta a_ev_ea_fv_f\Biggr]
\eeq

The angular distribution contains an additional $\cos\theta$
term, produced by the interference of the vector and the
axial current. 
This term induces a difference in the number of events
in the hemispheres on the electron and the positron side 
($\theta: (-\pi/2,\pi/2),~(\pi/2,3/2\pi)$ respectively).
This effect is called forward-backward asymmetry and
is generated by the $C/P$ violation of weak interactions.

\subsection{The Ratio $R$}

Integrating over the polar angle, we derive the total
QED cross section:
\beq 
\sigma~=~\frac{4\pi}{3}\alpha^2~e_f^2~\frac{1}{s}
\eeq

\noindent
Let us define the ratio
\beq
R~=~\frac{\sigma(e^+e^-\rightarrow hadrons)}
         {\sigma(e^+e^-\rightarrow \mu^+\mu^-)}
\eeq

\noindent
This way we get rid of uninteresting factors and concentrate 
on strong interactions only. On the experimental side, some
systematic effects, related to the luminosity determination for
example, cancel in taking the ratio. 

We assume now that the total hadronic cross section is
equal to the total partonic cross section, i.e.:
\beqn
\sigma(e^+e^-\rightarrow hadrons)&=&\sigma(e^+e^-\rightarrow partons)
\nonumber\\
&=&\sigma(e^+e^-\rightarrow q\overline{q})
+\sigma(e^+e^-\rightarrow q\overline{q}g)+...
\eeqn
There is a physical justification for that.
The time for the partonic process 
to occur (virtual photon decay) is
\beq
t_p~\sim~\frac{1}{E} 
\eeq
where $E$ is the center of mass energy,
while the time for hadronization is
\beq
t_h~\sim~\frac{1}{\Lambda_{QCD}}
\eeq
(or equivalently, the lifetime of a resonance 
$\tau\sim 10^{-23} sec$).
In high energy collisions, 
\beq
E~\gg~\Lambda_{QCD}, 
\eeq
hadronization occurs therefore much later than the partonic process, 
\beq
t_p~\ll~t_h,
\eeq
so its cross section is substantially unaffected.
The above qualitative explanation is not rigorous, but, as we will
see, its conclusions turn out to be rigth.
We have therefore in lowest order:
\beq
R~=~N_c~\sum_f e_f^2
\label{eq:tooeasy}\eeq
where the sum extends to all the quark flavors which are
kinematically 
allowed. The ratio $R$ has therefore the value:
\beq
R~=~N~(e_u^2+e_d^2+e_s^2)~=~3~(4/9+1/9+1/9)~=~2 
\eeq
below the charm-anticharm production threeshold, i.e. below
a center of mass energy of $\sim 3$ GeV, a value of
\beq
R~=~N~(e_u^2+e_d^2+e_s^2+e_c^2)~=~3~(4/9+1/9+1/9+4/9)~=~3.33 
\eeq
between the charm and the beauty production threshold, i.e.
between $\sim 4$ GeV and $\sim 10$ GeV, and
\beq
R=N(e_u^2+e_d^2+e_s^2+e_c^2+e_b^2)=3(4/9+1/9+1/9+4/9+1/9)=3.67 
\eeq
above the beauty threeshold and well below
the $Z$ peak (see fig. 5).
The above estimate is basically a counting of the degrees
of freedom involved.
It is remarkable that it depends only on the electric charges 
and the colour of the quarks.
The above prediction for $R$ is valid only far away from 
$c\overline{c}$ and $b\overline{b}$ resonances, where  
hadronization effects cannot be considered negligible
because they affect substantially the recombination of partons.
We see that $R$ in the resonance regions has peaks and valleys.
In these regions we may interpret the QCD estimate 
(\ref{eq:tooeasy}) as an average
over an interval of energy $\Delta E$ of order 1 GeV
\beq
R~\rightarrow~\langle R \rangle~=~\frac{1}{\Delta 
E}\int_{E-\Delta E/2}^{E+\Delta E/2} R(E')dE' 
\eeq
In the low energy range (a few GeV), the 
parton estimate is not very accurate because  hadron production
is still affected by light resonances.  
Furthermore, perturbative corrections (to be discussed later)
are large, and strongly dependent of unknown higher 
orders.

At $E=34~GeV$ the experimental
value of $R$ is about 3.9, to be compared to 3.67 (see fig.5).
The mismatch is only reduced by the $Z$ contribution
(it is a $O(1\%)$ effect, as can be evaluated inserting
the low energy approximation for $\chi_1$ in eq.(\ref{eq:basic})). 
As we will see, radiative corrections enhance the value
of the lowest order, bringing the theoretical value in
the experimental band. 

Let us consider now the resonance value ($E=M_Z$) of $R$.
Integrating over $\theta$ we have for the total cross section:
\beq
\sigma =\frac{4\pi\alpha^2 k^2}{3\Gamma_Z^2}
(a_e^2+v_e^2)(a_f^2+v_f^2)
\eeq
We note however that real photon emission from the initial 
$e^+e^-$ state
(called bremsstrahlung) smears the peak and shifts its position
(see fig. 6).
So, the previous formula has to be corrected for an accurate
quantitative analysis (see the end of sec. \ref{sud}
for a discussion of these effects). The ratio $R$ is given by:
\beq
R~=~3~\frac{\sum_{f=1}^5  a_f^2+v_f^2}{a_{\mu}^2+v_{\mu}^2}~=~20.09
\eeq
where we have used eqs.(\ref{eq:isospin}) 
and $\sin^2\theta_W = 0.2315$ \cite{pdg}.
We note that the QED effects mentioned
previously (bremsstrahlung) 
are the same for the muonic and hadronic
channels because  
they can be factorized as part of the initial
$e^+e^-$ annihilation process. 
So they cancel
in taking the ratio of the cross sections. 
The prediction
for the tree level value of $R$ on the peak therefore 
turns out to be accurate in this respect.
\noindent
Note that $R$ is much larger than in the QED case because 
quarks have comparable weak charges to that of the $\mu$. 

\noindent
The experimental value is \cite{alt}:
\beq
R_{exp}~=~20.80\pm 0.035
\eeq
Also in this case, the theoretical value is $\simeq 4\%$ smaller
than the experimental one, because of corrections to be
discussed in the next section.

\subsubsection{Radiative corrections}
\label{radcor}

We have seen in the previous section that the measures of $R$
are so accurate that a comparision with the theory requires
the inclusion of perturbative corrections.
Let us consider therefore radiative corrections to $R$.
It is a completely inclusive quantity and,
neglecting quark masses, it is characterized by a single energy scale,
\beq
\sqrt{s}
\eeq
(quark masses can be sent to zero because in this limit
no singularities are generated).
The order $\alpha_S$ corrections are ultraviolet finite because
the electromagnetic current which creates the $q\overline{q}$
pair is conserved in QCD (see the appendix).
In other words, the UV singularity of the vertex
correction is cancelled by the UV singularity of the
external quark legs renormalization. 
In the intermediate stage of the calculation,
infrared divergences appear, which cancel between real and
virtual diagrams in the completely inclusive process.
In this case they offer therefore only a technical problem.
Let us see this explicitely in dimensional regularization.
Infrared and collinear singularities (see later) are
regulated increasing the space-time dimension to $n>4$
(see for example \cite{sterman}).
The real corrections, with a gluon in the final state
(see fig. 7), give:
\beqn
\sigma(e^+e^-\rightarrow q\overline{q}g)
&=&\sigma_0\frac{C_F\alpha_S}{2\pi}~H(\epsilon)~
\int dx_1dx_2 \frac{x_1^2+x_2^2-\epsilon(2-x_1-x_2)}
{(1-x_1)^{1+\epsilon}(1-x_2)^{1+\epsilon}}
\nonumber\\
&=&\sigma_0\frac{C_F\alpha_S}{2\pi}~H(\epsilon)~\Bigg[
\frac{2}{\epsilon^2}+\frac{3}{\epsilon}+\frac{19}{2}+O(\epsilon)\Bigg]
\label{eq:reale}
\eeqn
where $C_F=\sum t_at_a=(N^2-1)/2N$ for an $SU(N)$ gauge theory
($C_F=4/3$ in QCD) is the Casimir operator in the fundamental 
representation, $n=4-2\epsilon$ and 
\beq
H(\epsilon)~=~\frac{3(1-\epsilon)^2}
{(3-2\epsilon)\Gamma(2-2\epsilon)}~=~1+O(\epsilon)
\eeq
The virtual corrections, i.e. the diagrams with a reabsorbed gluon, 
give:
\beq
\sigma(e^+e^-\rightarrow q\overline{q})
~=~\sigma_0\frac{C_F\alpha_S}{2\pi}~H(\epsilon)~\Bigg[
-\frac{2}{\epsilon^2}-\frac{3}{\epsilon}-8+O(\epsilon)\Bigg].
\label{eq:virtuale}
\eeq
Soft singularites appear as poles in $\epsilon$.
The double pole $1/\epsilon^2$ is originated 
by the product of the collinear
with the infared singularity (see later), while the simple poles 
$1/\epsilon$
come from a collinear {\it or} an infrared singularity.
Both double and simple poles cancel in the sum,
giving a finite $O(\alpha_S)$ correction to the tree level value of $R$:  
\beq
R~=~R_{tree}\Big(1+\frac{\alpha_S}{\pi}\Big)
\eeq
In the computation to order $\alpha_S^2$, we hit 
(no more cancelling) ultraviolet
divergences, of the form
\beq
\alpha_S^2\log\frac{\Lambda_0^2}{s},
\label{eq:uvbare}\eeq
where $\Lambda_0^2$ is an ultraviolet cutoff, i.e.
an ultraviolet regulator, and $\alpha_S=\alpha_S^{(0)}$ has to
be interpreted as the QCD bare coupling.
To be more explicit, the correction factor $K$ to $R$ up to $O(\alpha_S^2)$
has the form:
\beqn
K&=&1+C~\alpha_S-C~\frac{\beta_0}{4\pi}\alpha_S^2
~\log\frac{\Lambda_0^2}{s}~+C'\alpha_S^2+...
\nonumber\\
&=&1+C~\alpha_S\Bigl(1-\frac{\beta_0}{4\pi}~\alpha_S\log\frac{\Lambda_0^2}{s}
\Bigr)~+C'\alpha_S^2+...
\label{eq:unfactor}
\eeqn
(as we have seen $C=1/\pi$).
We can compare this result with the one-loop running coupling
of $QCD$ (eq.(\ref{eq:dopo})), in which we set $\mu^2=\Lambda_0^2$
and we interpret $\alpha_S(\Lambda_0^2)$ as the
bare coupling (we also set $Q^2=s$):
\beqn
\alpha_S(s)&=&\frac{\alpha_S(\Lambda_0^2)}
{1+\beta_0/(4\pi)\alpha_S(\Lambda_0^2)\log \Lambda_0^2/s}
\nonumber\\
&=&\alpha_S~\Biggl(1-
\frac{\beta_0}{4\pi}\alpha_S\log\frac{\Lambda_0^2}{s} +
\Biggl(\frac{\beta_0}{4\pi}\Biggr)^2\alpha_S^2
\log^2\frac{\Lambda_0^2}{s} +...\Biggr)
\label{eq:runcoup}
\eeqn
We may write, up to order $\alpha_S^2$:
\beq
K~=~1+C\alpha_S(s)+C'\alpha_S(s)^2+...
\label{eq:factor}
\eeq
We see that the logarithmic term in $K$
is simply a renormalization of the bare coupling.
The theory is trying to say that we are expanding
around a wrong scale, the ultraviolet cut-off
$\Lambda_0^2$, very far from the physical
scale $s$.
The unfactorized cross section (\ref{eq:unfactor}) 
and the factorized one (\ref{eq:factor})
coincide to order $\alpha_S^2$ included.
The terms added to generate the factorized form 
are indeed of higher order: $\alpha_S^3$ or more.

We see however that the above factorization is physically
reasonable but not unique. 
We may equally well write:
\beq
K~=~1+C\alpha_S-C\frac{\beta_0}{4\pi}~\alpha_S^2~\log\frac{\Lambda_0^2}{s/4}
~+~\Biggl[C'+C\frac{\beta_0}{4\pi}\log 4\Biggr]~\alpha_S^2+...
\label{eq:unfactor2}
\eeq
with a consequent factorized formula of the form:
\beq
K~=~1+C\alpha_S(s/4)+\Biggl[C'+C\frac{\beta_0}{4\pi}\log 4\Biggr]
\alpha_S(s/4)^2+... 
\label{eq:factor2}
\eeq
We see that the running coupling is evaluated at a different
scale, $s/4$, instead of $s$, and there is a compensating term
of order $\alpha_S^2$.
So, there is no way in perturbation theory to fix
the scale in the running coupling $\alpha_S$ to a unique value.
We can take whatever value for the scale, and compensate
shifting the finite $O(\alpha_S^2)$ term.
Of course, a wird scale setting of $10^{-4}s$ is unreasonable,
because it introduces a large logarithmic term $4\log 10\sim 10$
in the finite $\alpha_S^2$ term deteriorating the convergence
of the expansion.

The two factorized formulae (\ref{eq:factor}) and (\ref{eq:factor2})
agree formally up to order $\alpha_S^2$ but produce
slightly different numerical values.
They differ in the way they capture parts of the higher order
terms.
This phenomenon occurs because
of the different treatment of the coupling $\alpha_S$ and of $K$.
The running coupling resums classes of terms of
{\it any} order in $\alpha_S$, as it is clear from eq.({\ref{eq:runcoup}).
The group structure of scale tranformations (the renormalization
group) allows this resummation, while there is not any known
way to resum terms in $K$ to any order in $\alpha_S$:
$K$ is expanded up to a {\it fixed} order in $\alpha_S$, 
i.e. the series for $K$ is a truncated one.
When we change the scale of the coupling, an all-order
change, $K$ cannot compensate for the tail of the series.
The independence on the (arbitrary) scale choice
is therefore not rigorously true
in perturbation theory because it is violated by higher orders. 
This ambiguity never disappear: it can only be shifted to
higher orders as we compute more and more terms for $K$.
We have the formal independence from the scale choice of the
coupling only in the limit in which the infinite series
for $K$ has been computed.

Let us rephrase these facts in a more common language.
The UV divergence (\ref{eq:uvbare}) is converted with 
traditional renormalization in a similar logarithmic term:
\beq
\alpha_S^2\log \frac{s}{\mu^2},
\eeq
where $\mu$ is a renormalization point. 
In dimensional regularization (the only
regularization which is effectively used in high order
computations), after the 
subtraction of the $1/\epsilon$ poles, we are left
with logs of the above form, where $\mu$ is the mass
unit introduced to make the coupling dimensionless
in dimension $n\neq 4$.

$R$ is therefore written as a function of the
renormalized coupling at a fixed scale $\mu^2$, $\alpha_S(\mu^2)$,
and of the above logs:  
\beq
R~=~R(\alpha_S(\mu), \log s/\mu^2)
\eeq
since $R$ is a physical quantity, it cannot depend on the choice
of the renormalization scale $\mu^2$, which is an arbitrary scale
introduced for convenience at an intermediate stage (at the very end
we compare $R$ with another process characterized by another 
physical scale $s'$).
These considerations are formalized
with the renormalization group equation
\beq
\mu^2\frac{d}{d\mu^2} R~=~0,
\label{eq:rg}\eeq
where as usual, one has to vary $\mu^2$ remaining in the
same physical theory, i.e. keeping fixed
the bare parameters of the theory, or the renormalized parameters at
some other fixed scale $\mu_0^2$.
Explicitely eq.(\ref{eq:rg}) reads:
\beq
\Biggl[\mu^2\frac{\partial}{\partial\mu^2}
+\beta(\alpha_S)\frac{\partial}{\partial\alpha_S}\Biggr]
R(\alpha_S,\log s/\mu^2)~=~0
\label{eq:rg2}
\eeq
The concrete prediction we are able to give is 
slightly dependent on the renormalization point, an arbitrary scale
which has neither a physical meaning nor a definite value.
We need some physical intuition to pick up a 
good value of $\mu^2$ which minimize the (unknown) higher
orders: a natural choice seems to be 
$\mu^2=s$, for which $R$ is a (truncated) series
in $\alpha_S$ evaluated at the center of mass energy.
\beq
R~=~R(\alpha(s))  
\eeq
We have therefore:
\beq
R~=~K*R_{tree}
\eeq
where the correction factor (the so called $K$-factor) has
the following series expansion
\beq
K~=~1+\sum_{n=1}^{\infty} c_n \left(\frac{\alpha_S(s)}{\pi}\right)^n
\eeq
In the $\overline{MS}$ scheme (an off-shell renormalization
scheme, specific of dimensional regularization), three correction
terms are known \cite{chet,gorin}:
\beqn
c_1&=&1
\nonumber\\
c_2&=&\frac{365}{24}-11\zeta(3)+
     \Biggl(\frac{2}{3}\zeta(3)-\frac{11}{12}\Biggr))n_f
\nonumber\\
   &=&1.986-0.115~n_f~=~1.411
\nonumber\\
c_3&=&\frac{87029}{288}-\frac{1103}{4}\zeta(3)+\frac{275}{6}\zeta(5)
    -\frac{\pi^2}{48}\beta_0^2+\Biggl(\frac{55}{72}-\frac{5}{3}\zeta(3)
\Biggr)\eta
\nonumber\\
   &-&\Biggl(\frac{7847}{216}-\frac{262}{9}\zeta(3)+\frac{25}{9}\zeta(5)
\Biggr)n_f
    +\Biggl(\frac{151}{162}-\frac{19}{27}\zeta(3)\Biggl)n_f^2
\nonumber\\
   &=&-6.637-1.2n_f-0.005n_f^2-1.240 \eta
\nonumber\\
  &=&-12.80
\eeqn
where the last values of the coefficients are for $n_f=5$.
$\zeta(s)$ is the Rieman $\zeta$-function, defined by
the series
\beq
\zeta(u)~=~\sum_{n=1}^{\infty}\frac{1}{n^u}
\eeq
and
\beq
\eta~=~\frac{1}{3}\frac{(\sum_f e_f)^2}{\sum_f e_f^2}~=~\frac{1}{33}
~=~0.0303
\eeq
for the electromagnetic diagram (the last value is for $n_f=5$)
and
\beq
\eta~=~\frac{(\sum_f v_f)^2}{3\sum_{f}(v_f^2+a_f^2)}~=~0.0302
\eeq
for the weak diagram.
The different dependence on the electric/weak charges
of the $\eta$ terms is generated by diagrams with two separated
fermionic traces (like for example in the process 
$e^+e^-\rightarrow 3g$).

\noindent
At $\sqrt{s}=34$ GeV, taking $\alpha_s(34~GeV)=0.146\pm 0.030$,
we have $K\simeq 1.05$, i.e. the required $5\%$ increase of the
tree level value for $R$. 
At LEP1 energies, $\sqrt{s}=M_Z$, taking $\alpha_S(M_Z)=0.12$
\cite{alt},
we have $K\simeq 1.039$, i.e. a $4\%$ increase of the tree 
level value.
The experimental measure of $R$ 
is therefore a clean verification of QCD.
Note the large (negative) coefficient of the $\alpha_S^3$ term, which
makes the perturbative series not so well convergent.
For $\alpha_s=1$, for example, the $\alpha_S^3$ term is almost
three times bigger than the $\alpha_S^2$ term. 

The dependence of the coefficients $c_i$ on the scale,
shown explicitely in lowest order at the beginning of this
section, can be derived in general  
imposing the Callan Symanzik equation 
(\ref{eq:rg2}) order by order in $\alpha_S$.

\subsection{Jets in $e^+e^-$ collisions}

Perturbative QCD predicts also some properties of the structure of the final 
state. Of course, the cross section for a single exclusive channel,
like for example $e^+e^-\rightarrow P\overline{P}$, involves 
detailed  hadronization processes and is therefore
outside the reach of perturbative QCD.
We can consider seminclusive quantities.
Let us consider the emission of a gluon, i.e. the process
\beq
e^++e^-~\rightarrow~ q+\overline{q}+g
\eeq 
There are two amplitudes at order $\alpha_S$, 
for the emission of the gluon
from the quark leg $M_a$ or the antiquark one $M_b$ (fig.7).
The amplitudes involve only the QED-like  vertex
\beq
V_{q\overline{q}g}~=~ig\gamma_{\mu}t_a,
\eeq
i.e., the non abelian nature of QCD
(three and four gluon couplings) does not fully manifest itself.
The only effect of color is in the factor $C_F$ multiplying
$\alpha_S$. This implies that the QCD cross section is
identical to the abelian (QED) one with the replacement
\beq
C_F\alpha_S~\rightarrow~\alpha
\eeq
The non-abelian nature of QCD enters instead explicitely in the process
\beq
e^+e^-~\rightarrow~q\overline{q}gg,
\eeq
which is the dominant contribution to the 4-jet cross section
and starts at order $\alpha_S^2$. 
Let's go back to the 3-jet case.
We denote with $x_i$ the energy fractions of the quark, the
antiquark and the gluon:
\beq
x_1~=~\frac{E_q}{E_b},~~~~~x_2~=~\frac{E_{\overline{q}}}{E_b},~~~~~
x_3~=~\frac{E_g}{E_b},
\eeq
where $E_b$ is the beam energy ($\sqrt{s}=2 E_b$).
The conservation of energy and momentum gives:
\beqn
& &x_1+x_2+x_3~=~2
\nonumber\\
& &1-x_i~=~\frac{1}{2}x_jx_k(1-\cos\theta_{jk}),~~~~~i\neq j\neq k,~~~i\neq k.
\eeqn
The kinematical bounds are therefore
\beq
0\leq x_i\leq 1
\eeq
and
\beq
x_i+x_j\geq 1,~~~~i\neq j
\eeq
In words, a parton has the beam energy as the maximum energy
when the other two partons are collinear, and
a parton pair has the beam energy as the minimum energy
and the total energy as the maximum energy.

\noindent
Since the final state is planar, we have also:
\beq
\theta_{12}+\theta_{23}+\theta_{13}~=~2\pi
\eeq
The differential cross section 
for having a quark with an energy fraction $x_1$ and an antiquark
with an energy fraction $x_2$ in the final state is
\beq
\frac{1}{\sigma}
\frac{d\sigma_{q\overline{q}g}}{dx_1dx_2}~=~C_F\frac{\alpha_S}{2\pi}
\frac{x_1^2+x_2^2}{(1-x_1)(1-x_2)}
\label{eq:sqqg}\eeq
This cross section is derived in the appendix; it coincides
with the integrand in eq.(\ref{eq:reale}) 
in the limit $\epsilon\rightarrow 0$. 
It is symmetric under exchange of  $x_1$ with $x_2$, as it should be
for the $C/P$ symmetry of QCD
(to order $\alpha_S$, we
divide $d\sigma$ simply by $\sigma_0$).

If the experiment does not distinguish between the jets formed
by the quark, the antiquark and the gluon, it is natural
to symmetrize the above cross section: we simply ask what is
the cross section for having a jet with energy fraction $x_1$,
a second jet with energy fraction $x_2$ and a third jet with
energy fraction $x_3$. The cross section is therefore:
\beq
\frac{1}{\sigma}
\frac{d\sigma_{3jets}}{dx_1dx_2}~=~C_F\frac{\alpha_S}{6\pi}
\Biggl(\frac{x_1^2+x_2^2}{(1-x_1)(1-x_2)}
+\frac{x_1^2+x_3^2}{(1-x_1)(1-x_3)}
\frac{x_2^2+x_3^2}{(1-x_2)(1-x_3)}\Biggr)
\eeq
Since the jets are assumed to be undistinguishble, 
The energy fractions $x_i$ can be permuted so that
\beq
x_1<x_2<x_3
\eeq 
We therefore say that jet 1 is the less energetic one, jet 2
the intermediate one, and jet 3 the most energetic one.
Let's go back to the simpler formula (\ref{eq:sqqg}).
The integration region for the total cross section is
\beqn
0&\leq& x_1,x_2\leq 1
\nonumber\\
1&\leq& x_1+x_2
\eeqn
There are singularities when the quark
energy fractions reach their maximum allowed values 
(end point singularites):
\beq
x_i\rightarrow 1^-~~~~~~~~~~~i=1,2
\eeq
The strongest singularity is a double pole and occurs when
the energy fractions reach simultaneously their end point
values:
\beq
x_1~\rightarrow~1^-~~~~~~~~and~~~~~~~~x_2~\rightarrow~1^-
\label{eq:both}
\eeq
This happens when theee body kinematics 'resembles'
the two body one (the latter is a double spike in the
energy fractions $\sim\delta(1-x_1)\delta(1-x_2)$).

As we have seen, the double pole (\ref{eq:both}) 
is converted with the integration in dimensional
regularization in a double
pole $1/\epsilon^2$ (eq.(\ref{eq:reale})).
As we will see, the configuration (\ref{eq:both})
generates a double logarithmic contribution in 
the 3-jet cross section, i.e. a term of the form $\alpha_s\log^2y$.

\noindent
There is instead a simple pole singularity when only a 
single energy fraction $x_i$ reaches the singularity, 
\beq
x_i~\rightarrow~1^-,~~~~~~~~~x_j~<~1,~~~~~~~j\neq i
\label{eq:single}\eeq
(by $x_j<1$ we mean less than one by a finite amount,
$x_j=0.5$ for example).
This is explicitely seen decomposing the kernel as
\beq
\frac{x_1^2+x_2^2}{(1-x_1)(1-x_2)}~=~
\frac{2}{(1-x_1)(1-x_2)}-\frac{1+x_1}{1-x_2}-
\frac{1+x_2}{1-x_1}.
\eeq
The configuration (\ref{eq:single}) is related to simple 
$1/\epsilon$ poles
in dimensional regularization and to
single logarithmic terms
in the 3-jet cross section, i.e. to terms of the form
$\alpha_S\log y$.

Both the leading and the subleading singularities produce
divergences in the total cross section, so that
\beq
\sigma_{tot}(e^+e^-\rightarrow q\overline{q}g)=~\infty,
\label{eq:div}\eeq
which is an unphysical result.
Let us understand the origin of these singularites.
They originate technically from the quark
and antiquark propagators entering the Feynman amplitudes
(the kernel (\ref{eq:sqqg}) is roughly their product):
\beqn
\frac{1}{(p+k)^2}&=&\frac{1}{2 E_qE_g(1-\cos\theta_{qg})}
~=~ \frac{1}{s}~\frac{1}{1-x_2}
\nonumber\\
\frac{1}{(p'+k)^2}&=&\frac{1}{ 2E_{\overline{q}}E_g
(1-\cos\theta_{\overline{q}g}) }
~=~ \frac{1}{s}~\frac{1}{1-x_1} 
\eeqn
The propagators diverge for two different configurations:
\begin{enumerate}
\item[$i)$] 
Collinear singularity, $\theta\rightarrow 0$, i.e. when
the gluon is emitted at a very small ~~~angle with respect
to the source (the quark or the antiquark);
\item[$ii)$] 
Infrared singularity, $E_g\rightarrow 0$, i.e. when the
gluon is emitted with a very ~~~small energy.
\end{enumerate}
Let us see in detail how the end point singularites of (\ref{eq:sqqg})
are related to the infrared and the collinear ones.
The simple pole singularity
\beq
x_i~\rightarrow~1,~~~~x_j~<~1 
\eeq
implies
\beq
E_g>0,~~~~\theta_{j3}~\rightarrow~0,
\eeq
i.e. the collinear emission of a gluon with
a finite energy from the quark/ antiquark leg.
The simple pole singularities (\ref{eq:single}) are 
therefore induced by collinears but not soft gluons.

\noindent
Let us consider now the infrared limit
\beq
E_g~\rightarrow~0,~~~~~\theta_{qg}\neq 0,~~~~~\theta_{\overline{q}g}\neq 0
\eeq
i.e. a gluon with small energy but not collinear
to the quark or the antiquark.
Kinematics simplifies for
\beq
x_3~\ll~1
\eeq
to
\beqn
x_1&\simeq&1-x_3\frac{1-\cos\theta_{\overline{q}g}}{2}
\nonumber\\
x_2&\simeq&1-x_3\frac{1-\cos\theta_{qg}}{2}
\nonumber\\
\theta_{qg}&+&\theta_{\overline{q}g}~\simeq~\pi
\eeqn
such that
\beq
\cos\theta_{\overline{q}g}~\simeq~-\cos\theta_{qg}
\eeq
The infrared contribution to the total cross section is of the form
\beq
\delta\sigma~\sim~\alpha_S\int\frac{dx_1dx_2}{(1-x_1)(1-x_2)}~\sim~
\alpha_S\int\frac{dx_1dx_2}
{x_3^2 (1-\cos\theta_{qg})(1-\cos\theta_{\overline{q}g})}
\eeq
Introducing the new variables
\beqn
\epsilon&=&x_3
\nonumber\\
u&=&\cos\theta_{qg}
\eeqn
so that $x_1\simeq 1-\epsilon(1+u)/2,~x_2\simeq 1-\epsilon(1-u)/2$,
we have in the new variables
\beq
\delta\sigma~\sim~\alpha_S\int\frac{d\epsilon}{\epsilon}
                    \frac{du}{1-u^2}
\label{eq:IR}\eeq
Note that the change of variables converts the illusory
double pole in $x_3$ into a simple one.
Since $u\neq\pm 1$, we have a simple pole singularity
for $\epsilon\rightarrow 0$.
Infrared singularities are therefore associated with simple
poles.

We see therefore that both collinear and infrared singularities
are associated to simple poles of the differential cross section,
i.e. to single logarithmic terms in the 3-jet cross section
(see later).

Let us consider finally the case (\ref{eq:both}), i.e.
when the energy fractions $x_i$
reach simultaneously but independently their endpoint values.
In this case
\beq
E_g~\rightarrow~0~~~~~and~~~~~\theta_{qg}~\rightarrow~0~~~or~~~
                              \theta_{\overline{q}g}~\rightarrow~0
\eeq
Eq.(\ref{eq:IR}) is still valid, but in this case
\beq
\epsilon~\rightarrow~0~~~~~and~~~~~u~\rightarrow~\pm 1,
\eeq
giving in this case a double pole singularity.
We see that the leading singularity is generated by
a gluon which is soft and at the same time
collinear to the source
\beq
\delta\sigma~\sim~\alpha_S\int\frac{d\epsilon}{\epsilon}
                           \int\frac{d\theta}{\theta}        
\label{eq:easy}\eeq
where we have taken $u\simeq 1-\theta^2/2$ in eq.(\ref{eq:IR}).

We can integrate the differential cross section (\ref{eq:sqqg})
as long as we avoid the endpoints. 
The related cross section will be finite but strongly
dependent on the kinematical cuts.
We note that these cuts are naturally introduced in an
experiment in the form of resolution power.
Since detectors have a finite angular resolution $\delta>0$
we cannot disinguish an isolated quark from a quark
accompained by a gluon at an angle $\theta<\delta$.
Analogously, the finite energy resolution $\eta>0$ makes impossible
to distinguish a quark from a quark surrounded by a gluon with
energy $\epsilon<\eta$.
In other words, if the gluon is 'too' collinear and/or soft,
the final $q\overline{q}g$ is detected as a $q\overline{q}$
state:
\beq
\mid q\overline{q}g;~\epsilon<\eta,~\theta<\delta\rangle
~~\rightarrow~~\mid q\overline{q}\rangle_{phys}
\eeq
The divergent total cross section (\ref{eq:div}) is 
therefore unmeasurable.
The finite energy and angular 
resolution of the detectors
restrict the integration region of (\ref{eq:sqqg}) for the observable
$q\overline{q}g$ cross section, avoiding
the singular regions. Technically: there is a non zero lower limit
in the energy and angle integrations in (\ref{eq:easy}).
The observable cross section is therefore finite but dependent
on the resolution parameters $\delta$ and $\eta$.

\noindent
In the same spirit, let us 
consider now the {\it truly observable} cross section for
\beq
e^+e^-~\rightarrow~q\overline{q}
\eeq
to order $\alpha_S$.
The $q\overline{q}g$ cross section (\ref{eq:sqqg})
has to be integrated in the
singular region, i.e. in the region corresponding
to an undetectable gluon, but also virtual corrections to order 
$\alpha_S$ have to be included.
It turns out by explicit calculation that collinear 
and infrared singularites cancel in the sum, giving
rise to a finite cross section of order $\alpha_S$. 
That is basically the same cancellation which occurs
in the completely inclusive $O(\alpha_S)$ correction 
(eqs.(\ref{eq:reale}) and (\ref{eq:virtuale})), which include 
the same virtual diagrams and the same
singular integration regions for the real diagrams.

To summarize, the divergences encountered
mean that QCD has the tendency to
produce gluons with small energy and
with small opening angles with respect
to massless sources.

Collinear singularities (but not infrared ones) 
can be regulated with a small quark mass, which changes the
propagators according to:
\beqn
\frac{1}{(p+k)^2-m^2}&=&\frac{1}{2 E_qE_g(1-\beta_q\cos\theta_{qg})}
\nonumber\\
\frac{1}{(p'+k)^2-m^2}&=&\frac{1}{ 2E_{\overline{q}}E_g
(1-\beta_{\overline{q}}\cos\theta_{\overline{q}g}) }
\eeqn
where $\beta=p/E<1$ is the velocity. Therefore,
the gluons emitted in the production of beauty quarks,
i.e. in the process
\beq
e^+e^-~\rightarrow~b+\overline{b},
\eeq
do not have any collinear singularity.
The angular distribution of the gluons is given for small 
emission angles $\theta$ by
\beq
\frac{ dN}{d\cos\theta}~\sim~
\frac{1}{1-\beta\cos\theta}~\sim~\frac{2\gamma^2}{1+(\gamma\theta)^2}
\label{eq:angdib}\eeq
where $\gamma=E/m=1/\sqrt{1-\beta^2}\gg 1$ is the Lorentz factor.
The angular distribution has therefore a fast increase, 
\beq
\frac{ d N}{d\theta}~\sim~\theta\frac{dN}{d\cos\theta}~\sim
~\frac{1}{\theta}
\label{eq:masslessb}\eeq
when we reduce the emission angle from $\theta\sim 1$ up to
\beq
\theta_{min}~\sim~\frac{1}{\gamma}.
\eeq 
The behaviour (\ref{eq:masslessb}) is the same as that
of massless quarks. At even smaller angles
\beq
\theta~<\theta_{min}
\eeq 
the angular distribution is essentially constant.
That is how the quark mass $m$ acts as a collinear
regulator: it cutoffs the increase of the angular distribution. 
We expect therefore a logarithm of the mass after the
angular integration:
\beq
\int_{1/\gamma} \frac{d\theta}{\theta}~\sim~\log\frac{E}{m}
\eeq
The angular distribution (\ref{eq:angdib}) is similar to 
the angular distribution
of the classical electromagnetic radiation from
a relativistic charged particle (see \cite{jackson}).

The above discussion about physical states and observable cross
sections can be rephrased saying that 
we replaced single particle states with the true
{\it 'in'} and {\it 'out'} states, i.e.
the states prepared with accelerators and
observed with detectors.  
These are not anymore single particle states but
have components with different number of particles
\beq
\mid q\rangle_{phys}~=~a\mid q\rangle +b\mid q g\rangle
+c\mid q gg\rangle + d\mid qq\overline{q}\rangle+...
\eeq
The quark state is replaced by a quark surrounded
by a cloud of {\it collinear} and/or {\it soft} gluons and 
massless $q\overline{q}$ pairs.
In the next section we will give a simple prescription to build these
truly observable {\it out} states.

A remark is in order before ending this section. 
Up to now we considered the pure
perturbative process of gluon radiation, so any angular or energy cut for
the gluon is admissible (any $y$, see next section).
The only limitation to the above results 
{\it inside } perturbative QCD is related 
to the Landau pole: as we lower the momentum transfer
going into the soft region, the coupling rapidly increases
leaving the perturbative region.
But there is also a limitation of the applicability
of perturbative QCD results related to confinement.
A gluon with an energy of the order of the hadronic scale
\beq
\epsilon~\sim~\Lambda_{QCD}
\eeq
and/or with a transverse momentum $k_T$ with respect to the source
of the order of the hadronic scale,
\beq
k_T~\sim~\Lambda_{QCD}
\eeq
cannot move on a rectilinear motion neither producing a separate jet
(not enough energy)
nor flying away as an asymptotic state (confinement). 
This gluon is trapped by the color source.
The perturbative behaviour of such gluon (an almost free motion)
is therefore completely different from the 'real behaviour'.
If we want perturbative QCD to describe the real hadronic
world, we have to discard this conflict situation 
between perturbative and nonperturbative dynamics, imposing
energy and angular cuts well above the hadronic scale:
\beq
\epsilon~\gg~\Lambda_{QCD},~~~~~~~~k_T~\gg~\Lambda_{QCD}.
\eeq
That way we assign to different jets partons which can
effectively produce different jets.

\subsubsection{The Jade Algorithm}

The definition of an $N$ jet event according to the JADE algorithm
is the following \cite{jade}.
Take the 4-momenta of the particles $p_i$ and compute all the
invariant masses:
\beq
M_{ij}^2~=~(p_i+p_j)^2,~~~~i\neq j
\eeq
If the smallest invariant mass (of particles $i$ and $j$ say) is less than a 
given threshold,
\beq
M_{ij}^2~<~y~s,
\eeq
then particles $i$ and $j$ are combined together to form
a 'new' particle' (pseudoparticle) with momentum $p_{ij}=p_i+p_j$.
$y\ll 1$ is a constant determining how 'fats' are the jets,
such that $y s\gg \Lambda_{QCD}$ to avoid large hadronization 
effects. If instead
\beq
M_{ij}^2~\geq~y~s,
\eeq
particles $i$ and $j$ are not combined together.

We iterate this procedure treating pseudoparticles as particles. 
At each step two particles are combined into a psedupoparticle,
till the reduction stops because all the pair have
invariant masses greater than the threshold.
The final particles are identified with jets,
and their number is the jet multiplicity.

Experimentalists compute $n$ jet fractions combining 
the meaured momenta of the hadrons in the final state,
while theorists compuye $n$ jet fractions with 
final states composed of partons.
If the qualitative ideas about confinement and
hadronization discussed in secs.\ref {conf} and
\ref{jets} are correct, we
can test perturbative QCD comparing
hadronic jet fractions with partonic ones.

In the case of a $q\overline{q}g$ final state, a 3-jet
event is one in which all the three possible ivariant masses
exceed the threshold:
\beqn
(p+k)^2&\geq& ys\\
(p'+k)^2&\geq& ys\\
(p+p')^2&\geq& ys
\eeqn
which reads in terms of the energy fractions:
\beq
x_i \leq 1-y
\eeq
and therefore 
\beq
x_i+x_j\geq 1+y~~~~i\neq j
\eeq
We see that the boundary in the integration domain
(the dangerous terms) are eliminated.

\noindent
Let us define the jet fractions $f_i$ as
\beqn
f_2&=&\frac{\sigma_{2jet}}{\sigma_{tot}}
\nonumber\\
f_3&=&\frac{\sigma_{3jet}}{\sigma_{tot}}.
\eeqn
where $\sigma_{tot}=\sigma_{2jet}+\sigma_{3jet}=\sigma_0(1+\alpha_S/\pi)$ 
is the total 
cross section to order $\alpha_S$ (to compute $f_3$ to $O(\alpha_S)$
we divide simply by $\sigma_0$).
It holds clearly:
\beq
f_2+f_3~=~1.
\eeq
We have for $f_3$
\beqn
f_3&=&\int d\sigma_{q\overline{q}g}~=~C_F\frac{\alpha_S}{2\pi}
\int_{2y}^{1-y} \frac{dx_1}{1-x_1} \int_{1+y-x_1}^{1-y} 
\frac{dx_2(x_1^2+x_2^2)}{1-x_2}
\nonumber\\
&=&C_F\frac{\alpha_S}{2\pi}\Biggl(
4Li_2\left(\frac{y}{1-y}\right)+(3-6y)\log\left(\frac{y}{1-2y}\right)
+2\log^2\left(\frac{y}{1-y}\right)
\nonumber\\
& &~~~~~~~~~~~~~~~~-6y-\frac{9}{2}y^2-\frac{\pi^2}{3}+\frac{5}{2}\Biggr)
\label{eq:3jet}\eeqn
where $Li_2(z)$ is the dilogarithmic function (also called the Spence
function), defined by the integral representation:
\beq
Li_2(z)~=~-\int_0^z du~\frac{\log u}{1-u}
\eeq
We see that there are no infrared singularities 
in eq.(\ref{eq:3jet}) (like the 
$1/\epsilon$ poles we hit in eq.(\ref{eq:reale})):
the jet fraction is an 'infrared safe' quantity.
The jet fractions $f_3$ and $f_2=1-f_3$ are plotted in fig. 8
as a function of $y$ and in fig. 9 the experimental $f_i$
are represented.

If we vary the center of mass energy $\sqrt{s}$ of the
$e^+e^-$ collision keeping $y$ fixed, the jet fractions
$f_i$ change because of the scale dependence of the
coupling:
\beq
f_i(s,\alpha_S,y)~=~f_i(\alpha_S(s),y).
\eeq
We may therefore observe the variation of the QCD coupling
measuring jet fractions at different energy with the same
cut $y$. This is also a method for measuring $\alpha_S$
(see the lectures of B. Mele at this School).

\subsection{Sudakov Form Factors}
\label{sud}

Let us assume that we have an experiment with a good
resolution power in the measure of invariant masses,
i.e. that we can meaure very small invariant masses.
We can therefore measure the 2-jet fraction 
$f_2=1-f_3$ up to very small $y$
\beq
y~\ll~1
\label{eq:strong}\eeq
The condition (\ref{eq:strong}) is a 
strong restriction on the phase space
of the final hadrons/partons: the final
state consists of two very 'thin' jets in the space, which
contain almost all the energy of the event.

Neglecting terms which are infinitesimal with $y$
in eq.(\ref{eq:3jet})
(of the form $y,~y\log y,~y\log^2y,$ etc., the so called
power suppressed corrections), $f_2$ reads to order
$\alpha_S$:
\beq
f_2(y)~=~1-C_F\frac{\alpha_S}{2\pi}\Biggl(2\log^2 y+3\log y
           -\frac{\pi^2}{3}+\frac{5}{2}+O(y)\Biggr)
\label{eq:f21}\eeq
There are three terms.
There is a double logarithmic term which is the remnant of the
collinear times infrared singularity, a single
logarithmic term which is the remnant of the infrared
$+$ collinear singularity, and a finite term. Simbolically:
\beq
f_2^{(1)}~\sim~\alpha_S\log^2y+\alpha_S\log y +\alpha_S
\eeq
The $y$ cut, as discussed previously, regulates both kind of
singularites.
If we use instead 
two different regulators for collinear and
infrared singularities (for example an energy and an angular cut),
we have for $f_2$ an expression of the form:
\beq
f_2^{(1)}~\sim~\alpha_S Ll+\alpha_SL+\alpha_S l+\alpha_s
\eeq
where $L$ is the collinear logarithm 
(containing the angle cut) and $l$ the infared 
logarithm (containing the energy resolutiom).

\noindent
The $O(\alpha_S^2)$ corrections to $f_2$ have also been computed
\cite{ali}:
\beq
f_2^{(2)}~=~\Bigl(\frac{\alpha_S}{2\pi}\Bigr)^2
\Biggl(C_F^2Z_C(y)+C_FNZ_N(y)+T_RZ_T(y)\Biggr)
\label{eq:f22}\eeq
where:
\beqn 
Z_C(y)&=&2\log^4y+6\log^3y+\Big(\frac{13}{2}-\zeta(2)\Big)\log^2y
+\Big(\frac{9}{4}-3\zeta(2)-12\zeta(3)\Big)\log y
\nonumber\\
&&~~~~~~~~~~~~+\frac{1}{8}-\frac{51}{4}\zeta(2)+11\zeta(3)+4\zeta(4)
\nonumber\\
Z_N(y)&=&\frac{11}{3}\log^3y+\Big(2\zeta(2)-\frac{169}{36}\Big)\log^2y
+\Big(6\zeta(3)-\frac{57}{4}\Big)\log y
\nonumber\\ 
&&~~~~~~~~~~~+\frac{31}{9}
+\frac{32}{3}\zeta(2)-13\zeta(3)+\frac{45}{4}\zeta(4)
\nonumber\\
Z_T(y)&=&-\frac{4}{3}\log^3 y+\frac{11}{9}\log^2 y+5\log y
+\frac{19}{9}-\frac{38}{9}\zeta(2)
\label{eq:f22def}\eeqn
We see that the $O(\alpha_S^2)$ correction to $f_2$ contains
up to four logarithms of $y$. In general 
the correction to order $\alpha_S^n$ contains terms up
to $\log^{2n}y$, i.e. there at most two logarithms per loop
(one collinear times one infrared log for each loop).
We have therefore:
\beq
{f_2}^{(n)}~=~\alpha_S^n\sum_{k=0}^{2n} C^{(n)}_k\log^k y
~=~C^{(n)}_{2n}\alpha_S^n\log^{2n}y
+C^{(n)}_{2n-1}\alpha_S^n\log^{2n-1}y+...
\eeq
The perturbative
expansion for $f_2(y)$ is not simply an expansion in
powers of $\alpha_S$ at the relevant scale, like in the case of
$R$.
Every order in $\alpha_S$ is a polynomial in $\log y$.
That means that there are logarithms which are not absorbable in
the renormalization of the coupling constant, as it
does happen instead with $R$.
Let us consider the physical difference between these
two observable.
$R$ is an inclusive quantity,
characterized by a single mass scale, the center of mass
energy $\sqrt{s}$, so the theory can develop
only logarithms of the form 
\beq
\log\frac{\Lambda_0^2}{s},
\eeq
where $\Lambda_0^2$ is the ultraviolet cutoff,
which are absorbed by renormalization of the 
lagrangian since the theory is renormalizable.
The jet fraction $f_2$ is instead a seminclusive quantity and its
definition involves also another scale, $ys$.
As a result, the theory can develop (and it actually does)
logs of the other possible mass ratio,
\beq
\log\frac{s}{ys}~=~\log\frac{1}{y}
\label{eq:ly}\eeq 
These logs are {\it final}, i.e. they do not cancel in the final
result, as it does happen with the virtual and real diagram
infrared singularities in $R$.
The logs (\ref{eq:ly}) have not ultraviolet origin 
(they are indeed infrared, as we saw),
so they are not absorbable by a coupling redefiniton.
 
For $y\ll 1$ these logarithms can become so large 
\beq
\log\frac{1}{y}~\gg~1
\eeq
that
\beq
\alpha_S\log^2 y~\sim~1
\eeq
even though
\beq
\alpha_S~\ll~1
\eeq
The convergence of the expansion
in powers of $\alpha_S$ may therefore be spoiled by such large
logarithmic coefficients.
The solution of this problem is to abandon fixed order
perturbation theory, in favour of an expansion
according to the {\it degree of singularity}  
of the terms in the limit
\beq
\log \frac{1}{y}~\rightarrow~\infty.
\eeq
The most singular terms, corresponding to the lowest order
in this new asymptotic expansion, are of the form
\beq
\alpha_S^n\log^{2n}y~~~~~~~~~n=0,1,2,3,...,k,....
\label{eq:morsin}\eeq
The sub-leading terms are of the form
\label{eq:lead}\beq
\alpha_S^{n}\log^{2n-1}y~~~~~~~~~n=1,2,3...k...
\eeq
and are of order $1/\log y\ll 1$ smaller than the leading ones.
The sub-sub-leading terms are of the  form
\beq
\alpha_S^{n}\log^{2n-2}y~~~~~~~~~n=1,2,3...k...
\eeq
and are of order $1/\log y$ smaller than the sub-leading ones
and so on. 
Keeping only the leading series if often called 
double logarithmic approximation (DLA).
Let us see how the two series are related.
By simple rearrangement:
\beqn
& &f_2~=~
\nonumber\\
&=&1+a_0\alpha x^2+a_1\alpha x+a_2\alpha
+b_0\alpha^2 x^4+b_1\alpha^2 x^3+b_2\alpha^2x^2
+b_3\alpha^2x+b_4\alpha^2+...
\nonumber\\
&=&\Big(1+a_0\alpha x^2+b_0\alpha^2x^4+...\Big)
+\Big(a_1\alpha x+b_1\alpha^2x^3+...\Big)
+\Big(a_2\alpha +b_2\alpha^2 x^2+...\Big)
\nonumber\\
&+&\Big(b_3\alpha^2 x+...\Big)+...
\eeqn
where $x=\log y$, 
a more confortable notation for the coefficients has
been used and we dropped the subscript on $\alpha_S$.

Since the coefficients of the leading terms (\ref{eq:morsin})
are known for any order (they exponentiate, see later),
we can extend the validity of our result toward smaller
$y$ values factorizing the leading series.
Up to order order $\alpha^2$ we have
\beqn
f_2&=&( 1+a_0\alpha x^2+b_0\alpha^2x^4+...\Big)
\Big(1+a_1\alpha x+a_2\alpha +
(b_1-a_0a_1)\alpha^2x^3+
\nonumber\\
& &(b_2-a_0a_2)\alpha^2x^2
+b_3\alpha^2x+b_4\alpha^2+...\Big)
\eeqn
We factorized the leading terms subtracting the spurious
terms in the remaining factor up to $O(\alpha^2)$ included.
The difference with the unfactorized expression is therefore
$O(\alpha^3)$.
In the first bracket on the right hand side we can
add all the higher order terms of the leading series,
i.e. resume the whole leading series,
without affecting the second bracket
up to order $\alpha^2$:
\beqn
f_2&=&( 1+a_0\alpha 
x^2+b_0\alpha^2x^4+c_0\alpha^3x^6+...+f_0\alpha^n x^{2n}+...\Big)
\Big(1+a_1\alpha x+a_2\alpha 
\nonumber\\
&+&(b_1-a_0a_1)\alpha^2x^3
+(b_2-a_0a_2)\alpha^2x^2
+b_3\alpha^2x+b_4\alpha^2+...\Big)
\eeqn
We may factorize the subleading series in the same way:
\beq
f_2=(lead.)\Big(1+a_1\alpha x+b_1'\alpha^2x^3+...\Big)
\Big(1+a_2\alpha+b_2'\alpha^2x^2+(b_3-a_1a_2)\alpha^2 x
+b_4\alpha^2\Big)
\eeq
where $b_1'=b_1-a_0a_1,~b_2'=b_2-a_0a_2$ are the new
coefficients shifted by the factorization of the
leading series. This procedure can be carried on 
to any prescribed order in $\alpha_S$.

Note that the coefficient of the $\log^4 y$ term in ({\ref{eq:f22})
is 1/2 of that of the $\log^2 y$ term in (\ref{eq:f21}):
\beq
f_2~=~1-C_F\frac{\alpha_S}{\pi}\log^2 y +\frac{1}{2}
\Biggl(\frac{C_F\alpha_S}{\pi}\Biggr)^2\log^4 y+...
\eeq
It can be proved that the leading terms of any order do exponentiate,
i.e. they give a contribution to $f_2$ of the form

\beq
f_2~=~e^{-\alpha_S C_F/\pi\log^2y}
\label{eq:sud}
\eeq

Expressions of this form are known as Sudakov form factors,
who studied similar problems in QED long ago 
\cite{collins,kodaira,landau,trentadue}.
Since in lowest order $f_2=1$, we see that the leading 
double logarithmic terms suppress the 2-jet cross section
for $y\ll 1$.
There is a physical explanation for that.
For $y\ll 1$ we are considering the quasielastic production
of $q\overline{q}$ pairs at high energy, i.e. a final
state with little activity around the tree level produced
$q\overline{q}$ pair.
Accelerated color
charges naturally produce radiation, as it happens in 
classical electrodynamics.
The Sudakov represents the suppression of the improbable 
non radiative channels.
We see here an implementation of the
factorization idea discussed in sec. \ref{pm}: 
the basic electromagnetic 
process is corrected by soft gluon effects which appear 
as a factor in eq.(\ref{eq:sud}).  
The important physical information given by the Sudakov
form factors is how the measured cross sections depend
on the cut and the resolution of the experiments.

Another important property of the Sudakov form factors
is the broadening of the sharp structures, like for example
peaks in cross sections around resonances.
Let us consider a QED case, the $Z$ line shape discussed previously.
The Sudakov suppression of non radiative channels
implies a relative increase of the radiative cross sections,
so that energy is frequently released from the $e^+e^-$
system to the radiation field,
through multiple photon emissions:
\beq
e^+e^-~\rightarrow~e^+e^-+n\gamma
\eeq
The fluctuations in the energy released by the $e^+e^-$
pair induce equal fluctuations in the energy available
for the production of the resonance, so the latter is not anymore
produced always on the peak, for any selected beam energy.
That smears the peak and shifts its position toward
higher center of mass energies $\sqrt{s}>M_Z$.

Let us observe that also the subleading terms proportional
to $(C_F)^n$, i.e. the QED-like ones, seem to exponentiate:
\beq
f_2~=~1-\frac{C_F\alpha_S}{2\pi}
\Big(2\log^2y+3\log y\Big)+\frac{1}{2}
\Biggl(\frac{C_F\alpha_S}{2\pi}\Biggr)^2
\Big(4\log^2 y +12\log^3 y\Big)+...
\eeq
They do indeed exponentiate,
producing the correction factor

\beq
\delta f_2~=~e^{-3\alpha_SC_F/(2\pi)\log y}.
\label{eq:abel}\eeq

The terms (\ref{eq:abel}) however do not contain the whole
subleading corrections, as is clear looking to the
form of the $Z_N(y)$  and $Z_T(y)$
contributions in (\ref{eq:f22def}) (they both contain
$\alpha_S^2\log^3 y$ terms). 
The factorization of all the subleading corrections is a complicated
problem because it involves many different phenomena:
the variation of $\alpha_S$ with the scale
(if for example $\alpha_S(\mu)\rightarrow 
\alpha_S(y\mu)+O(\alpha_S^2\log y)$, 
$\delta(\alpha_S\log^2y)=O(\alpha_S^2\log^3y)$, i.e. a subleading term),
separate collinear and infrared singularities.
  
\section{Conclusions}

In these lectures we discussed the foundation of perturbative
QCD and an important physical application, the high-energy
$e^+e^-$ annihilation into hadrons.
Even though perturbative QCD is technically similar
to the perturbative expansion of QED, the physical
content is very different.
In QED the physical one-particle states (electrons 
and photons) are identical to the ones in the free theory
(free electrons and photons) after renormalization. 
One then makes a perturbative expansion of the scattering 
matrix elements, so only a (weak) convergence assumption
of the series is postulated.
In perturbative QCD the one-particle states (the hadrons)
are not identical at all to the one-particle states in the free
theory (free quarks and gluons), after renormalization.
We neglect this problem and compute cross
sections with quarks and gluons {\it as} they were asymptotic states.  
We relate subsequenty these {\it unphysical} processes
to the {\it physical} ones with a set of qualitative 
ideas about their connection (parton-hadron duality).
Apart from convergence problems, perturbative QCD computations 
therefore have to be supplemented by assumptions about non-perturbative 
phenomena (such as confinement, hadronization, jet formation),
necessarily accompaining the perturbative process.
In these lectures
we discussed this point considerably,
presenting also qualitative discussions 
and some models of confinement and jet formation.

In the second part of the paper we considered the $e^+e^-$
annihilation to hadrons at high energy, which has many
different dynamical aspects.
The simplest observable is the total hadronic cross section.
Its perturbative computation relies only
on the assumption that hadronization does not change 
the probability of the partonic process.
Furthermore, the perturbative expansion is very simple:
it involves only numerical coefficients times power
of $\alpha_S$ evaluated at the center of mass energy.
More 'delicate' observables are the jet fractions. 
Their perturbative computation requires an extra
assumption with respect to the total cross section.
We have to assume local parton-hadron duality,
i.e. that hadrons are formed by partons which
are close in phace space.
This postulate is necessary because
jet fractions contain dynamical effects 
related to intermediate scales, i.e. momenta $p$
between a cutoff scale $\Lambda$ ($\gg\Lambda_{QCD}$)
and the center of mass energy 
\beq
\Lambda^2~<~p^2~<s.
\eeq
These scales have not to be mixed by hadronization.
Also the perturbative series for the jet fractions
is more complicated: there are large logarithmic
corrections of the form
\beq
\log\frac{s}{\Lambda^2}.
\eeq 
The latter is anyway a technical problem that is solved
{\it within} perturbative QCD with resummation techniques.

Comparing with experimental data, we have seen that the
perturbative QCD approach,
perturbative computations + hadronization assumptions,
{\it does indeed work,} both qualitatively and quantitatively.
A face of the hadronic world seems to be largely understood.
This also implies that confinement is very 'hidden'
in many processes.
The progress in the perturbative QCD direction consists in 
higher order computations, better resummation techniques, and in finding
measurable quantities that characterize some 
properties of the hadronic states and 
are little influenced by hadronization.

\appendix
\section{Evaluation of the 3-jet cross section}

In this section we compute the cross section at order $\alpha_S$
for
\beq
e^+e^-~\rightarrow~q\overline{q}g
\eeq
We take the gluon propagator in Feynman gauge:
\beq
S^{\mu\nu}_{ab}(k)~=~\frac{-ig^{\mu\nu}\delta_{ab}}{k^2+i\epsilon},
\eeq
and the quarks and leptons massless.
The Feynman amplitude $M$ is:
\beqn
& & M~=~M_a+M_b=
\\
&-&ie^2gt_a\overline{v}_{r'}(l')\gamma_{\mu}u_{r}(l)
\frac{g_{\mu\nu}}{s+i\epsilon}
\overline{u}_s(p)
\Bigg(\gamma_{\rho}\frac{ \hat{p}+\hat{k} }{(p+k)^2}\gamma_{\nu}-
\gamma_{\nu}\frac{ \hat{p'}+\hat{k} }{(p'+k)^2}\gamma_{\rho}\Bigg)
v_{s'}(p')\epsilon^{\rho}_{ah}(k)
\nonumber\eeqn
where $\hat{p}=\gamma_{\mu}p^{\mu}$,
$s=q^2$, $q=l+l'$ and $l$ and $l'$ are the 4-momenta
of the electron and the positron.

\noindent
We take now the square of the modulus, average over the initial
helicities and sum over the final helicities and
polarizations, 
\beq
\frac{1}{4}\sum_{rr'ss'h}\mid M\mid^2,
\eeq
using the formulas
\beqn
\sum_r u_r(p)\overline{u}_r(p)&=&\hat{p},~~~~~~~~
\sum_r v_r(p')\overline{v}_r(p')~=~\hat{p'}
\nonumber\\
\sum_h\epsilon_{h\mu}(k)\epsilon_{h\nu}(k)&=&-g_{\mu\nu}
             +\frac{k_{\mu}k_{\nu}}{k^2} 
\eeqn
(the $k_{\mu}k_{\nu}/k^2$ term in the photon polarization sum
does not contribute because the diagrams are QED-like).

\noindent
We multiply by the relativistic final states factor
\beq
\frac{d^3 p}{(2\pi)^3 2p_0}~
\frac{d^3 p'}{(2\pi)^3 2{p'}_0}~
\frac{d^3 k}{(2\pi)^3 2k_0},
\eeq
the relativistic normalization factor for the initial state
\beq
\frac{1}{2l_0}\frac{1}{2l_0'}~=~\frac{1}{4E_b^2},
\eeq
the 4-momentum conserving $\delta$ function
\beq
(2\pi)^4\delta^{(4)}(q-p-p'-k),
\eeq
and divide by the relative velocity $v_{rel}$ of the $e^+e^-$ pair in the 
center of mass frame (i.e. by the flux for volume $V=1$):
\beq
v_{rel}~=~\frac{ l\cdot l'}{l_0l_0'}~=~2,
\eeq
(the two particles have opposite light velocity). We have:
\beqn
d\sigma&=&
\frac{1}{2^{13}\pi^5E_b^2}
\sum_{rr'ss'h}\mid M\mid^2
\delta^{(4)}(q-p-p'-k)
\frac{d^3 p}{p_0}
\frac{d^3 p'}{ {p'}_0}
\frac{d^3 k}{ k_0}
\nonumber\\
&=&-\frac{C_F e^4 g^2}{2^{13} \pi^5}~\frac{1}{E_b^2 s^2}~
L_{\mu\nu}(l,q)~T^{\mu\nu}(q)
\eeqn
where the leptonic and the (completely integrated)
hadronic tensors are given by
\beqn
L_{\mu\nu}(q,l)&=&Tr[\hat{l}\gamma_{\mu}\hat{l}'\gamma_{\nu}]
\nonumber\\
T_{\mu\nu}(q)&=&\int~
\frac{d^3p}{p_0}\frac{d^3p'}{{p'}_0}\frac{d^3 k}{k_0}~
\delta^{(4)}(p+p'+k-q)~
H_{\mu\nu}(p,p',k)
\eeqn
and $H_{\mu\nu}$ is the (unintegrated) hadronic tensor
\beq
H_{\mu\nu}=
Tr\Biggl[
\hat{p}\Biggl(\gamma_{\rho}\frac{\hat{p}+\hat{k}}{(p+k)^2}\gamma_{\mu}-
\gamma_{\mu}\frac{\hat{p'}+\hat{k}}{(p'+k)^2}\gamma_{\rho}\Biggr)
\hat{p'}\Biggl(\gamma_{\nu}\frac{\hat{p}+\hat{k}}{(p+k)^2}\gamma^{\rho}
-\gamma^{\rho}\frac{\hat{p'}+\hat{k}}{(p'+k)^2}\gamma_{\nu}\Biggr)
\Biggr]
\eeq
The tensors are symmetric under exchange of the indices $\mu$ and $\nu$
and are transverse with respect to $q_{\mu}$ 
because of electromagnetic current conservation
\beq
q_{\mu}~T^{\mu\nu}~=~0,~~~~~~~~q_{\mu}~L^{\mu\nu}~=~0,
\eeq
$T_{\mu\nu}$ can be parametrized as
\beq
T_{\mu\nu}(q)~=~(g_{\mu\nu}-\frac{q_{\mu}q_{\nu}}{q^2})~T(q^2)
\eeq
so that
\beq
T(q^2)~=~\frac{1}{3}g^{\mu\nu}T_{\mu\nu}(q).
\eeq
The contraction gives
\beq
L_{\mu\nu}~T^{\mu\nu}~=~\frac{1}{3}g^{\mu\nu}L_{\mu\nu}~
         g^{\rho\sigma}T_{\rho\sigma}
~=~-\frac{4}{3}~q^2~g^{\rho\sigma}T_{\rho\sigma}.
\eeq
We decoupled the tensors with the projector
$g_{\mu\nu}g_{\rho\sigma}/3$. 
Therefore we need only to compute the contraction 
\beqn
&& g^{\mu\nu}~H_{\mu\nu}(p,p',k)~=
\\
&&Tr\Biggl[
\hat{p}\Biggl(\gamma_{\rho}\frac{\hat{q}-\hat{p}'}{(q-p')^2}\gamma_{\mu}-
\gamma_{\mu}\frac{\hat{q}-\hat{p}}{(q-p)^2}\gamma_{\rho}\Biggr)
\hat{p'}\Biggl(\gamma^{\mu}\frac{\hat{q}-\hat{p}'}{(q-p')^2}\gamma^{\rho}
-\gamma^{\rho}\frac{\hat{q}-\hat{p}}{(q-p)^2}\gamma^{\mu}\Biggr)
\Biggr]
\nonumber\eeqn
Because of the $\delta^{(4)}(p+p'+k-q)$ we made the replacements
\beq
p+k~=~q-p',~~~~~p'+k~=~q-p.
\eeq
We have:
\beqn
g^{\mu\nu}~H_{\mu\nu}(p,p',q-p-p')&=&
\frac{X_{11}}{((q-p')^2)^2}
+\frac{X_{12}}{(q-p)^2(q-p')^2}+
\nonumber\\
&+&\frac{X_{21}}{(q-p)^2(q-p')^2}
+\frac{X_{22}}{((q-p)^2)^2}
\eeqn
where
\beqn
X_{11}&=&+Tr\Biggl[
\hat{p}\gamma_{\rho}(\hat{q}-\hat{p}')\gamma_{\mu}
\hat{p'}\gamma^{\mu}(\hat{q}-\hat{p}')\gamma^{\rho}
\Biggr]
\nonumber\\
X_{12}&=&-Tr\Biggl[
\hat{p}\gamma_{\rho}(\hat{q}-\hat{p}')\gamma_{\mu}
\hat{p'}\gamma^{\rho}(\hat{q}-\hat{p})\gamma^{\mu}
\Biggr]
\nonumber\\
X_{21}&=&-Tr\Biggl[
\hat{p}\gamma_{\mu}(\hat{q}-\hat{p})\gamma_{\rho}
\hat{p'}\gamma^{\mu}(\hat{q}-\hat{p}')\gamma^{\rho}
\Biggr]
\nonumber\\
X_{22}&=&+Tr\Biggl[
\hat{p}\gamma_{\mu}(\hat{q}-\hat{p})\gamma_{\rho}
\hat{p'}\gamma^{\rho}(\hat{q}-\hat{p})\gamma^{\mu}
\Biggr]
\eeqn
The explicit computation gives:
\beqn
X_{11}&=&4 
Tr\Big[\hat{p}~(\hat{q}-\hat{p}')~\hat{p'}~(\hat{q}-\hat{p}')\Big]
~=~4~Tr\Big[\hat{p}~\hat{q}~\hat{p'}~\hat{q}\Big]
\nonumber\\
&=&16\Big(2p\cdot q ~p'\cdot q-q^2~p\cdot p'\Big)
\nonumber\\
X_{12}&=&
2~Tr\Big[
\hat{p}\gamma_{\rho}(\hat{q}-\hat{p}')(\hat{q}-\hat{p})
\gamma^{\rho}\hat{p'}\Big]
=8(q-p)\cdot(q-p')~Tr\Big[p\cdot p'\Big]
\nonumber\\
&=&32~p\cdot p'~(q-p)\cdot (q-p') 
\nonumber\\
X_{21}&=&X_{12}
\nonumber\\
X_{22}&=&X_{11}
\eeqn
where we used the identities
\beqn
&&\gamma_{\mu}\gamma_{\rho}\gamma^{\mu}=-2\gamma_{\rho}
\nonumber\\
&&Tr\Big[\gamma_{\mu}\gamma_{\nu}\gamma_{\rho}\gamma_{\sigma}\Big]
=4\Big(g_{\mu\nu}g_{\rho\sigma}+g_{\mu\sigma}g_{\nu\rho}
      -g_{\nu\sigma}g_{\mu\rho}\Big)
\nonumber\\
&&\gamma_{\mu}\gamma_{\alpha}\gamma_{\beta}\gamma_{\rho}\gamma^{\mu}
=-2\gamma_{\rho}\gamma_{\beta}\gamma_{\alpha}.
\eeqn
Summing all the terms we derive:
\beqn
g^{\mu\nu}~H_{\mu\nu}&=&
16~\Big(2p\cdot q ~p'\cdot q-q^2p\cdot p'\Big)
\Biggl[\frac{1}{((q-p)^2)^2}+\frac{1}{((q'-p)^2)^2}\Biggr]
\nonumber\\
&+&64 ~\frac{p\cdot p' (q-p)\cdot(q-p')}{(q-p)^2(q-p')^2}, 
\label{eq:hadtr}
\eeqn
The scalar projection of $H_{\mu\nu}$ is much simpler 
to compute than the original tensor because it
is easily reduced to the trace 
of four gamma matrices, while the computation
of $H_{\mu\nu}$ involves the trace of six gamma matrices. 

We perform now the integrations in the center of mass frame,
where
\beq
\vec{q}~=~0.
\eeq
First we integrate 
the $\delta^{(3)}(\vec{p}+\vec{p}'+\vec{k})$
over the gluon 3-momentum $\vec{k}$, which gives
\beq
\vec{k}~=~-\vec{p}-\vec{p}',
\eeq
and therefore
\beq
k_0~=~\mid \vec{p}+\vec{p}'\mid ~=~ \sqrt
{ p_0^2+{p'}_0^2+2p_0{p'}_0\cos\theta_{q\overline{q}} }.
\eeq
We choose now the spatial frame in such a way that $\vec{p}$
is directed along the $z$-axis and integrate over the polar
angle of $\vec{p}'$,
\beq
\frac{d^3 p'}{p_0'}~=~p_0'dp_0'd\cos\theta' d\phi',
\eeq
The energy-conserving $\delta(k_0+p_0+p_0'-q_0)$-function gives:
\beq
\int \frac{p_0dp_0d\Omega p_0'dp_0'd\phi'd\cos\theta'}{k_0}
\delta(q_0-p_0-p_0'-k_0)~g_{\mu\nu}H^{\mu\nu}=
\int dp_0d\Omega dp_0'd\phi'~g_{\mu\nu}H^{\mu\nu}
\eeq
because
\beq
\Biggl[\frac{\partial k_0}{\partial \cos\theta'}\Biggr]^{-1}
~=~\frac{k_0}{p_0p_0'}.
\eeq
The integration over $d\Omega d\phi'$ is trivial and gives $8\pi^2$:
\beq
\int dp_0dp_0'd\Omega d\phi'g_{\mu\nu}H^{\mu\nu}
~=~8\pi^2E_b^2 dx_1 dx_2 g_{\mu\nu}H^{\mu\nu}
\eeq
We express now the hadronic trace (\ref{eq:hadtr}) in terms of energy 
fractions $x_1$ and $x_2$. We divide numerators and denominators
by $E_b^4$, and replace the rescaled scalar products according to
the relations
\beq
p\cdot q~=~2x_1,~~~~p\cdot q'~=~2x_2,~~~~q^2~=~4,~~~~
p\cdot p'~=~2(x_1+x_2-1).
\eeq
These substitutions give:
\beqn
g^{\mu\nu}H_{\mu\nu}&=&8\Biggl[\frac{1-x_1}{1-x_2}+\frac{1-x_2}{1-x_1}+
\frac{2(x_1+x_2-1)}{(1-x_1)(1-x_2)}\Bigg]
\nonumber\\
&=&8\frac{x_1^2+x_2^2}{(1-x_1)(1-x_2)}
\eeqn
We note that the leading singularity $(1-x_1)^{-1}(1-x_2)^{-1}$
in Feynman gauge comes from the interference term
(while it comes from a single direct term in a particularly choosen
axial gauge \cite{russi,germ}).

Putting all the pieces together, we have the final result:
\beq
d\sigma~=~\frac{2C_F}{3}\frac{\alpha^2\alpha_S}{s}
d x_1d x_2\frac{x_1^2+x_2^2}{(1-x_1)(1-x_2)}
\eeq 
which coincides with result (\ref{eq:sqqg}) multiplied by 
$\sigma_0=4\pi/3\alpha^2/s$.
Performing the angular integrations and the
Dirac algebra in $n$ dimensions, we derive the dimensionally
regularized cross section in eq.(\ref{eq:reale}). 

\vfill
\eject

\newpage

\tableofcontents

\newpage

\centerline{\bf FIGURE CAPTIONS}

\begin{enumerate}
\item[Fig.1:] A two jet event from hadronic $Z$ decay (DELPHI,
taken from ref. \cite{pich}).
\item[Fig.2:] A three jet event from hadronic $Z$ decay (DELPHI,
taken from ref. \cite{pich}).
\item[Fig.3:] Born diagrams for $e^+e^-\rightarrow q\overline{q}$.
\item[Fig.4:] Angular distribution of two jet events at 34 GeV
center of mass energy. 
The curve is the QED + parton model prediction $1+x^2$,
where $x=\cos\theta$ (TASSO, from ref. \cite{salsiv}).
\item[Fig.5:] Compilation of $R$ values at low energy (upper figure)
and at high energy (lower figure), (from ref. \cite{pdg}).
\item[Fig.6:] Total cross section for 
$e^+e^-\rightarrow\mu^+\mu^-$ +e.m. radiation,
around the $Z$ peak. The dotted line is the Born approximation,
the dashed line is the Born approximation + leading logs
coming from initial state radiation and the solid line is
the Born approximation + leading logs + subleading logs
(from ref. \cite{salsiv}).
\item[Fig.7:] Diagrams for real corrections (upper figure)
and virtual corrections (lower figure) of order $\alpha_S$
to $e^+e^-\rightarrow q\overline{q}$.
\item[Fig.8:] Plot of the jet fractions $f_2$ and $f_3$ to order   
$\alpha_S$ (from ref. \cite{kell}).
\item[Fig.9:] QCD fits to the jet fractions (OPAL, from ref. \cite{kell}).
\end{enumerate}


\begin{thebibliography}{999} 

\bibitem{me}
U. Aglietti  and Z. Ligeti, Phys. Lett. B 364 (1995) 75-77.

\bibitem{ali}
A. Ali, {\it QCD $\gamma\gamma$ and heavy quark physics},
in PHYSICS AT LEP, Vol.2 (Cern Yellow Reports) CERN 86-02 (1986),
pag.99; see also the references therein to the original articles.

\bibitem{alt} 
G. Altarelli, Cern preprint CERN-TH-95/196.
(July 1995);

\bibitem{alt2}
G. Altarelli, Phys. Rep 81, n.1 (1982) 1-129.

\bibitem{ash}
N.W. Ashcroft, N.D. Mermin, {\it Solid State Physics,}
HRW International Editions (1976), pag. 731.

\bibitem{jade}  
S. Bethke et al. (JADE Collaboration), Phys. Lett. 213B 
(1988) 235; for improved algorithms, see N. Brown and
W.J. Stirling, Durham preprint, RAL-91-049 DTP/91/30  (June 1991).

\bibitem{beta3l}
W.E. Caswell, Phys. Rev. Lett 33 (1974) 244;
D.R.T. Jones, Nucl. Phys. B75 (1974) 531;
O.V. Tarasov, A.A. Vladimirov and A. Yu Zharkov,
Phys. Lett 93B (1980) 429.

\bibitem{cl}
T.P. Cheng e L.F. Li, {\it Gauge Theory of Elementary Particles},
Clarendon Press (1984), pag. 280.

\bibitem{chet} % R a 3 loop
K.G. Chetyrkin, A.L. Kataev and F.V. Tkachov,
Phys. Lett. 85B (1979) 277;
M. Dine and J. Sapirstein, Phys. Rev. Lett. 43 (1979) 668;
W. Celmaster and R. Gosalves, Phys. Rev. Lett 44 (1980) 560.

\bibitem{collins}
J.C. Collins, {\it Sudakov Form Factors}, in {\it Perturbative
Quantum Chromodynamics}, ed. A.H. Mueller (World Scientific,
Singapore, 1989) 573-614. 

\bibitem{russi}
Yu.L. Dokshitzer et al., {\it Basics of Perturbative QCD,}
Editions Frontieres (1991).

\bibitem{kell} 
R.K. Ellis, {\it QCD at Tasy '94}, Fermilab preprint
FERMILAB-Conf-94/410-T December 9, 1994.

\bibitem{frigel}
H. Fritzsch and M. Gell-Mann in: Proc XVIth Inter. Conf. on
High Energy Physics (Chicago-Batavia 1972, Vol.2 pag. 135;
H. Fritzsch, M. Gell-Mann and H. Leutwyler, Phys. Lett. B47
(1973) 365.

\bibitem{gorin}
S.G. Gorishny, A.L. Kataev and S.A. Larin,
Phys. Lett. B259 (1991) 144;
L.R. Surguladze and M.A. Samuel, Phys. Rev. Lett 66 (1991) 560.

\bibitem{gw}
K. Gottfried and V.F. Weisskopf, 
{\it Concepts in Particle Physics}, Oxford University Press (1988);
for a discussion on the antiscreeninig of QCD vacuum pag.380,
and for confinement pag. 397.

\bibitem{germ}
W. Greiner and A. Schafer,{\it Quantum Chromodynamics}, Springer (1994);
this textbook contains many computations described in great detail
(it is a german book), among which there is the calculation of the 
3-jet cross section in axial gauge, pag. 292.

\bibitem{jackson}
J.D. Jackson, {\it Classical Electrodynamics,} 
John Wiley, New York 1975.

\bibitem{kodaira}
J. Kodaira and L. Trentadue, Phys. Lett. 112 B, n.1 (1982) 66-70;
L. Trentadue, SLAC-PUB-2934 June 1982 (T).

\bibitem{landau}
L.D.  Landau and E.M. Lifsits, {\it Teoria Quantistica Relativistica,}
Fisica Teorica 4, Editori Riuntiti Edizioni MIR (1978)
({\it Quantum Field Theory}, Vol. IV, A Course in Theoretical Physics).

\bibitem{tdlee} 
T.D. Lee, {\it Particle Physics and 
Introduction to Field Theory}, Harwood Academic Publishers (1988).

\bibitem{marti}
G. Martinelli,{\it Perturbative QCD}, Proceedings of the 
1987 JINR-CERN School of High Energy Physics 
(near Varna, Bulgaria, 6-19 september 1987).

\bibitem{nason}
P. Nason, M. Magnoli and R. Rattazzi, UPRF-90-292 preprint,
October 1990; there is a clear discussion about the assumptions
on hadronization on which perturbative QCD relies.

\bibitem{pdg}
Particle Data Group, {\it Review of Particle Physics}, Phys. Rev. D54 (1996).

\bibitem{pich}
A. Pich, {\it Quantum Chromodynamics}, lectures given at the 
1994 European School
of High Energy Physics (Sorrento, Italy, 29 August - 11 September 1995).

\bibitem{asfr}
H.D. Politzer, Phys. Rev. Lett. 30 (1973) 1346;
D.J. Gross and F. Wilczek, Phys. Rev. Lett. 30 (1973) 1343.

\bibitem{salsiv}
G. Salvini and A. Silverman, Phys. Rep. 171 (n.5 and 6) (1988).

\bibitem{sterman}
G. Sterman, {\it An Introduction to Quantum Field Theory},
Cambridge University Press (1993).

\bibitem{trentadue}
L. Trentadue, Phys. Lett. 151 B, n.2 (1985), 171-173.

\bibitem{nonabel}
C.N. Yang and F. Mills, Phys. Rev. 96, 191 (1954).

\bibitem{qm} 
See for example, A. Le Yaouanc et al., {\it Hadron Transitions in the
Quark Model}, Gordon and Breach Science Publishers (1988), pag.44.

\bibitem{wilson}
K. Wilson, Phys. Rev. 179 (1969) 1499. %OPE 

\end{thebibliography}
\end{document}